\documentclass[12pt]{article}
\usepackage{multirow}
\usepackage{graphics}
\usepackage{lscape}
\usepackage{amsfonts}
\usepackage{amsmath}
\usepackage{amssymb}
\usepackage{rotating}
\usepackage{amssymb,amsmath}
\usepackage[usenames]{color}
\usepackage{graphicx}
\usepackage[english]{babel}
\usepackage{natbib}
\usepackage{array}
\newcolumntype{P}[1]{>{\centering\arraybackslash}p{#1}}
\usepackage{array,multirow}
\usepackage{hhline}

\usepackage{float}
\usepackage{xcolor}
\usepackage{ulem} % This makes the text strike out.
\usepackage{actuarialangle}
\usepackage[flushleft]{threeparttable}
\usepackage{comment}
\makeatletter

\newcommand{\Rmnum}[1]{\expandafter\@slowromancap\romannumeral #1@}
\makeatother
\usepackage{etoolbox}
\appto\TPTnoteSettings{\footnotesize}
\renewcommand{\section}[1]{\refstepcounter{section}\begin{center}{\sc \thesection. #1}\end{center}}
\renewcommand{\subsection}[1]{\bigskip \noindent \refstepcounter{subsection}{\thesubsection\ {\it #1}}}
\renewcommand{\subsubsection}[1]{\bigskip \noindent \refstepcounter{subsubsection}{\thesubsubsection\ {\it #1}}}
\renewcommand{\title}[1]{{\bf \begin{center}#1\end{center}}}
\renewcommand{\author}[1]{\begin{center}{\sc #1}\end{center}}

\newcommand{\acknowledgements}{\begin{center}{\sc Acknowledgements}\end{center}}
\newcommand{\competinginterests}{\begin{center}{\sc Competing Interests}\end{center}}
\renewenvironment{abstract}{\begin{center}{\sc abstract}\end{center}\small}{\normalsize}
\newenvironment{keywords}{\begin{center}{\sc keywords}\end{center}\small}{\normalsize}

\newenvironment{contact}{\begin{center}{\sc contact address}\end{center}\small}{\normalsize}

\newenvironment{bajlist}{\begin{list}{(\alph{bajlistnum})\hfill}{\usecounter{bajlistnum}\setlength{\labelwidth}{0.3in}\setlength{\leftmargin}{0.3in}\setlength{\rightmargin}{0in}\setlength{\labelsep}{0in}\setlength{\topsep}{0in}\setlength{\partopsep}{0in}\setlength{\itemsep}{0in}\setlength{\parsep}{0in}}}{\end{list}{}\bigskip}
\newenvironment{bajsublist}{\begin{list}{(\arabic{bajsublistnum})\hfill}{\usecounter{bajsublistnum}\setlength{\labelwidth}{0.3in}\setlength{\leftmargin}{0.3in}\setlength{\rightmargin}{0in}\setlength{\labelsep}{0in}\setlength{\topsep}{0in}\setlength{\partopsep}{0in}\setlength{\itemsep}{0in}\setlength{\parsep}{0in}}}{\end{list}{}}

\newtheorem{proposition}{Proposition}
\newtheorem{corollary}{Corollary}

\def\ordd[#1][#2,#3;#4,#5]{\renewcommand{\arraystretch}{0.2}% 
                            \setlength{\arraycolsep}{0pt}%
                            \begin{array}{ccc}%
                                 \scriptstyle#2 &  & \scriptstyle#3\\%
                                 \scriptstyle#4 &#1& \scriptstyle#5%
                           \end{array}}

\newcommand{\Prob}{\mbox{P}}
\newcommand{\Mean}{\mbox{E}}

\newcounter{bajlistnum}
\newcounter{bajsublistnum}

\begin{document}

\newcommand{\HRule}{\rule{\linewidth}{0.5mm}} % Defines a new command for the horizontal lines, change thickness here

%----------------------------------------------------------------------------------------
%	TITLE SECTION
%----------------------------------------------------------------------------------------

\title{A REVIEW OF THE MARKOV MODEL OF  LIFE INSURANCE \\  WITH A VIEW TO SURPLUS}

\author{By Oytun Ha\c{c}ar{\i}z\dag, Torsten Kleinow\ddag $\,$ and  Angus S. Macdonald\S}

\begin{abstract}

{We review Markov models of surplus in life insurance based on a counting process  following \cite{norberg1991}, uniting probabilistic theory with elements of practice largely drawn from UK experience. First, we organize models systematically based on one and two technical bases, including a suitable descriptive notation. Extending this to three technical bases to accommodate different valuation approaches leads us: (a) to expand the definition of `technical basis’ to include non-contractual cashflows recognized in the associated Thiele equation; and (b) to add new (mainly) systematic terms to the surplus. Making these cashflows dynamic or `quasi-contractual' covers many real applications, and we give two as examples, the paid-up valuation principle and reversionary bonus on participating contracts.}

\end{abstract}

%-------------------------------------------------------------

\begin{keywords}

\noindent Actuarial Basis, Life Insurance, Loadings, Surplus, Technical Basis

\end{keywords}

\begin{contact}

\noindent \dag $\,$ Department of Actuarial Sciences, The Faculty of Business, Karab{\"{u}}k University, Karab{\"{u}}k, 78050, Turkey, and Institute of Applied Mathematics, Middle East Technical University, Ankara, 06800, Turkey.

\noindent \ddag $\,$ Research Centre for Longevity Risk, Faculty of Economics and Business, University of Amsterdam.

\noindent \S $\,$ Department of Actuarial Mathematics and Statistics, Heriot-Watt University, Edinburgh EH14 4AS, UK, and the Maxwell Institute for Mathematical Sciences, UK.

\noindent Corresponding author: Angus Macdonald, A.S.Macdonald@hw.ac.uk.

\end{contact}

\section{Introduction}
\label{sec:Intro}

\subsection{Motivation: From Northampton to Copenhagen}
\label{sec:MotivationI}

In 1989, Hans B\"uhlmann designated life insurance actuaries as `actuaries of the first kind' and called them `essentially deterministic' \citep{buehlmann1989}. The motto of the Institute of Actuaries, {\em certum ex incertis}, tells us why; it means `certainty out of uncertainty', and life insurance actuaries were descended from 17th and 18th century probabilists whose key idea was expectation rather than probability \citep{daston2023}. Their main tools for managing risk were: (a) safe-side margins in premiums and reserves, and: (b) retrospective distribution of any surpluses that emerged. The  technical basis used to calculate premiums would build {loadings} into the rates to be charged, and the technical basis used to calculate reserves then had to stop loadings in future premiums from emerging as surplus too soon. A natural and {passive} approach was to use the same technical basis for both premiums and reserves, known as the `Northampton method' after the life table originally used by the Equitable Life in the UK \citep{fisher1965}. 

%----------------------------------------------------------------

However, life insurance contracts have very long terms indeed. In time, the original technical basis may become archaic, so that using it to calculate the liability seems absurd, raising serious doubts about the balance sheet. During the 19th century, the Northampton method was abandoned by UK companies, in favour of more active valuations on more up-to-date technical bases. Debates on methodology enlivened the early years of the UK actuarial profession \citep{turnbull2017}. In time for the age of computers, profit-testing emerged \citep{anderson1959},  in which future loadings were calculated and valued individually and explicitly.

A first step in moving actuaries `of the first kind' beyond deterministic thinking was to express life table probabilities such as ${}_tp_x$ and actuarial present values such as $\bar{A}_x$ as respectively $\Prob[X]$ and $\Mean[Y]$, for well-defined random variables $X$ and $Y$ (the latter involving discounting as well as mortality). Beginning with   \cite{hickman1964}, the future lifetime of the canonical `individual alive at age $x$', named $(x)$, was defined to be a random variable $T_x$. In the course of only a few generations this has become the usual basis for educating student actuaries \citep{bowers1986, gerber1990, dickson2020}. So {${}_tp_x = \Prob[ T_x > t ]$}, the present value of a whole-life benefit of \pounds 1 payable immediately on the death of $(x)$ is $v^{T_x}$, and by definition $\bar{A}_x = \Mean[ v^{T_x}]$. 

%-------------------------------------------------------------

\markright{A Review of the Markov Model of Life Insurance With a View to Surplus}

%-------------------------------------------------------------

The landmark papers \cite{hoem1969, hoem1988} introduced actuaries to multiple-state Markov models. The advance from an alive-dead mortality model to multiple states perhaps obscured the advance in thinking represented by modelling a {process}. A focus on transition probabilities and intensities (often rendered as ${}_tp_{x}^{ij}$ and $\mu_{x+t}^{ij}$ in homage to tradition) invited again the question, `what is the underlying random object?' The answer was in \cite{hoem1978}, which defined cashflows contingent upon the {counting process} $N(t)$, counting the number of deaths up to and including time $t$. The seeming triviality of $N(t)$, which takes only the values 0 or 1, may have concealed its virtues. Based on $N(t)$, \cite{aalen1978} launched a revolution in survival modelling, see \cite{andersen1993} and  \cite{aalen2022}.

\cite{buehlmann1976} had suggested modelling the surplus under a life insurance contract using a suitable martingale. With $N(t)$ (in its multivariate version) representing underlying stochastic events, \cite{ramlau-hansen1988a, ramlau-hansen1988b} carried this through in great generality for Markov models. \cite{norberg1991} explored the definitions of reserves, prospective and retrospective, in terms of conditional expectations with respect to quite general {streams of information, not only filtrations\footnote{In this work we assume that information is represented by filtrations. See \cite{norberg1991} for an example in a multiple-state setting where this is not the case.}.  This work essentially identified technical bases as probability measures.} Extensions to handle the distribution of surplus as bonus were added in \cite{ramlau-hansen1991} and \cite{norberg1999, norberg2001}.  

Much of this work by what may justly be called the `Scandinavian school' was summarized in the encyclopedia article \cite{norberg2004b}. However, all of it assumes what we call `Scandinavian-style regulation'. That is, premiums and reserves are calculated on the same safe-side `first order' technical basis, and the experience is represented by another `second order' technical basis. The reserve at outset is zero, and surplus emerges as the difference between the two technical bases. In other words, the Northampton method. 

{Thus we have two lines of approach, even two traditions: the long history and variety of practice that provoked B\"uhlmann to define `actuaries of the first kind'; and in a more limited setting, the stochastic process approach of the Scandinavian school. Most students today are shown a half-way house between the two, namely life insurance mathematics in terms of the random lifetime $T_x$.}

%---------------------------------------------------------------------------------------------

\subsection{Objectives}
\label{sec:Objective}

\begin{bajlist}

\item As well as \cite{norberg2004b}, excellent graduate-level books have been written (\cite{moeller2007}, \cite{koller2012}, \cite{asmussen2020}) but the basic {counting process} framework is still scattered in the literature. \cite{norberg2004b} {noted that: ``[an]} account of counting processes and martingale techniques can be compiled from \cite{hoem1978} and \cite{norberg1999}'' and {doing} that is where we start. 

\item However, Scandinavian-style regulation confines margins (loadings) to the elements of the technical basis, which is too limiting. To describe more realistic regimes we need to extend valuation technical bases to include hypothetical policy cashflows. The main practical effect is to define loadings in the premiums, and to control when these emerge in the surplus. 

\item Further, some practical applications require policy cashflows to be modelled dynamically, in which case we call them quasi-contractual. We finish with two such examples; the paid-up valuation principle of \cite{linnemann2002}, and reversionary bonus.

\end{bajlist}

{Mention of the stochastic present value $v^{T_x}$ in Section \ref{sec:MotivationI} invites two questions. One is  how to stochasticize the discount factor $v$? This would lead us into financial economics.  The other is how to generalize the biometric model represented by $T_x$? That would lead us to the multivariate version of $N(t)$, the unstated stochastic object underlying \cite{hoem1969, hoem1988}. Along such lines, \cite{norberg1992} essentially recast life insurance mathematics into the same form as no-arbitrage financial mathematics. Both questions are subsidiary to our particular aim and to try to address them would simply add a great deal of notation, so apart from a few remarks we leave them aside.}

%-------------------------------------------------------------

\subsection{Plan of the Paper}
\label{sec:Plan}

In Section \ref{sec:Elementary} we start with an elementary model of a balance sheet to fix ideas. Section  \ref{sec:NotationsRepresentations} defines notation for cashflows based on stochastic representations of a lifetime. We leave aside multiple-state models and stochastic assets to focus on the structure of surplus; these extensions follow along clear lines laid down  in the literature. Section \ref{sec:TechBasesOverall} defines technical bases in terms of Thiele's differential equation, and the key identification of policy values both as solutions of Thiele's equation and as conditional expectations.  Section \ref{sec:BalanceSheet} defines assets and liabilities in the model balance sheet, leading to the definition of surplus, then in Sections \ref{sec:modelingBalanceSheet} and \ref{sec:StochAndmodeled} we introduce surplus based respectively on one or two technical bases, including a systematic notation. Thus far we review material sketched in \cite{norberg2004b} up to and including Scandinavian-style regulation.  Section \ref{sec:DiffPremValBases} separates premium and valuation technical bases and defines both pure and valuation premiums and the associated loadings as in \cite{hacariz2024}. Section \ref{sec:TechBasisCashflows} generalizes this by defining an actuarial basis to be a technical basis and associated policy cashflows assumed in the associated Thiele equation. In Section \ref{sec:ModifiedDecomp} we extend the analysis of surplus to three technical bases, adding a loading term to the systematic component of surplus. To handle cases in practice where the actual policy cashflows are only quasi-contractual, subject to options at the choice of one party, Section \ref{sec:ChangingValuationBasis} introduces a time-indexed family of  actuarial bases, and we present two examples, the paid-up valuation principle in Section \ref{sec:PUPValnPrincipleMain}, and reversionary bonus in Section \ref{sec:Bonus}. Conclusions are in Section \ref{sec:Concs}.

%-------------------------------------------------------------
%-------------------------------------------------------------
%-------------------------------------------------------------

\section{An Elementary Model of a Life Insurer's Balance Sheet}
\label{sec:Elementary}

\subsection{A Deterministic Model}
\label{sec:Deterministic}

To fix ideas, consider a {deterministic model of a} life insurance policy {with term $n$ years, commencing at time $t=0$ on an individual aged $x$, who is hereafter referred to as $(x)$.} A sum insured of $S$ is paid immediately on death at time $t$, and a maturity benefit of $\bar{S}$ (possibly zero) is paid on survival for $n$ years. Level premiums are paid throughout the term at rate $P$ per year. We  ignore expenses.

The technical basis consists of a force of interest $\delta_t$ and force of mortality $\mu_{x+t}$ (which we denote by $\mu_t$, taking age $x$ as understood) at time $t \ge 0$. Define $\varphi(t)$ to be a discount factor allowing for survivorship as well as interest:

\begin{equation}
\varphi(t) = \exp \left( - \int_0^t ( \delta_r + \mu_{r}) \, dr \right). \label{eq:IntegratingFactor}
\end{equation}

\noindent The premium rate $P$ is the solution of the equation of value:

\begin{equation}
0 = \int_0^n \varphi(r) \, (P - \mu_{r} \, S) \, dr - \varphi(n) \, \bar{S}. \label{eq:EqPrin}
\end{equation}

\begin{figure}
\begin{center}
\includegraphics[scale=0.87]{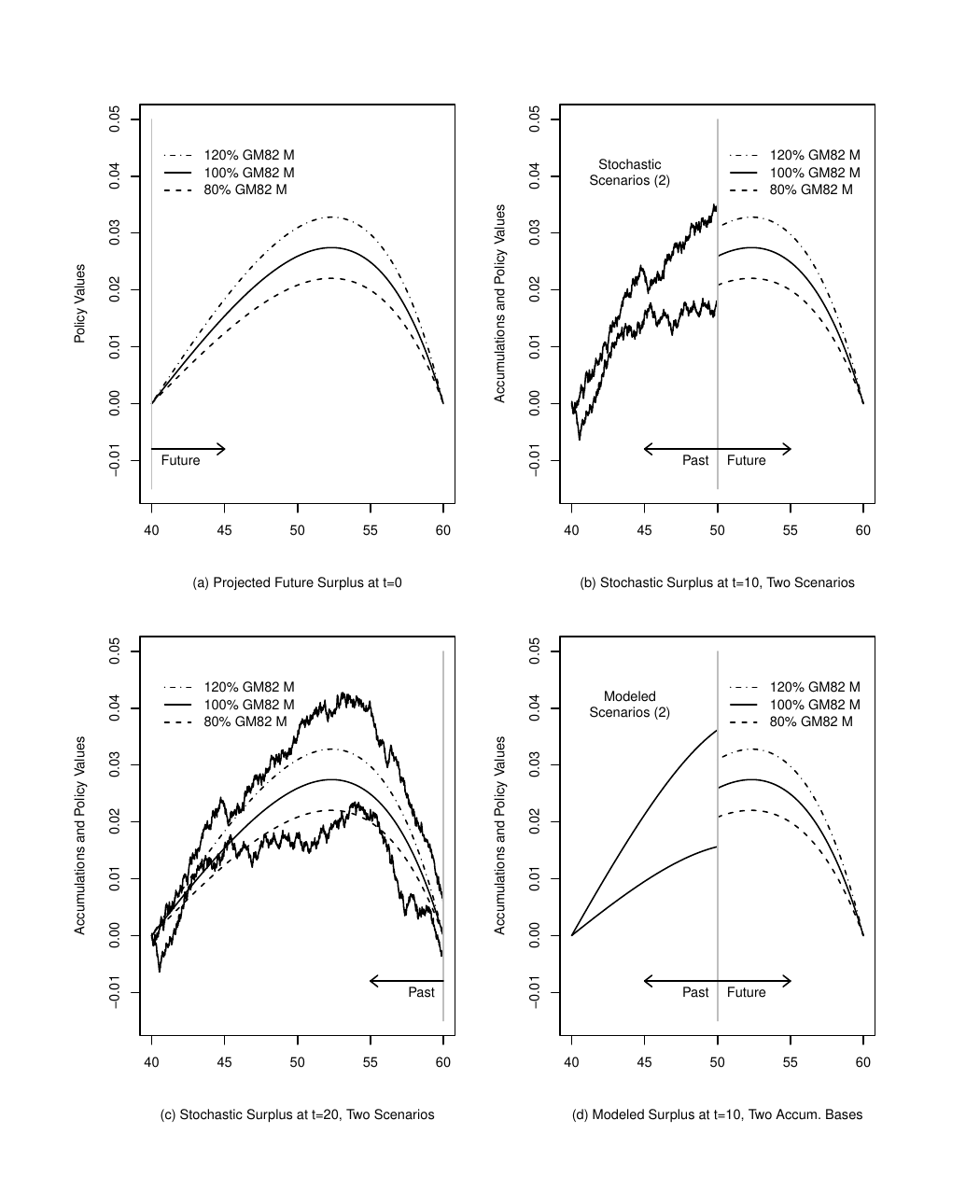}
\caption{\label{fig:TechBases} The time evolution of balance sheets. {Large portfolio of term insurances}, sum assured \$1, age 40, term 20 years. Baseline {technical basis:} force of interest $\delta=0.05$ and mortality $\mu_t$ = GM82 Males (Denmark). (a) projected policy values at $t=0$, baseline and two other technical bases; (b) stochastic surpluses at $t=10$, two scenarios of accumulations, projected policy values on three technical bases; (c) stochastic surpluses at $t=20$, two scenarios of accumulations, past policy values on three technical bases for comparison; (d) modeled surpluses at $t=10$, two  accumulation bases up to $t=10$.}
\end{center}
\end{figure}

\noindent Dividing both sides by $\varphi(t)$ and splitting the integral at time $t$ $(0 \le t \le n)$ we have:

\begin{equation}
0 = \underbrace{\int_0^t \frac{\varphi(r)}{\varphi(t)} \, ( P - \mu_{r} \, S ) \, dr}_{\mbox{\scriptsize Retrospective Policy Value = $W_t$}} - \underbrace{\int_t^n \frac{\varphi(r)}{\varphi(t)} \, ( \mu_{r} \, S - P ) \, dr - \frac{\varphi(n)}{\varphi(t)} \, \bar{S}}_{\mbox{\scriptsize Prospective Policy Value = $V_t$}} \label{eq:EqPrinPrem1.5}
\end{equation}

% Equation (\ref{eq:EqPrinPrem1.5}) is a mathematical model of the insurer's balance sheet at time $t$, and we call an equation of this kind a {\em balance sheet equation}. 

\noindent in which we identify {the assets, or  retrospective policy value, denoted by $W_t$, and the liabilities}, or prospective policy value, denoted by $V_t$, equal at all times. Both are `per policy in-force at time $t$', and are solutions of {Thiele's differential equation}:

\begin{equation}
\frac{d}{dt} \, g(t) = \delta_t \, g(t) + P - \mu_t \, ( S - g(t) ) \label{eq:ThieleElementary}
\end{equation}

\noindent satisfying the respective boundary conditions $g(0)=0$ and  $g(n)=\bar{S}$. {Surplus is defined as $W_t-V_t$ and is uniformly zero.} This model is illustrated by panel (a) of Figure \ref{fig:TechBases}, which shows the smooth progression of term insurance policy values with $x=40, n=20, (\bar{S}=0)$ on three different technical bases\footnote{The contract is a 20-year term assurance for a man age 40, and the technical basis for both (level) premium rate and policy values (referred to as the first-order technical basis) is interest of $\delta=0.05$ {\em per annum} and the mortality of the Danish GM82 Males life table, with no expenses.}.

%-------------------------------------------------------------

%--------------------------------------------------------------

\subsection{Valuation Functions}
\label{sec:ValuationFunctions}

The valuation problem may now be stated. Let $\mathbf{H} = \{ {\cal H}_t \}_{t \ge 0}$ represent a flow of information {(filtration)} available at times $t\ge 0$. We need to choose appropriate asset and liability valuation functions $W_t^*({\cal H}_t)$ and $V_t^*({\cal H}_t)$ to replace $W_t$ and $V_t$ from Section \ref{sec:Deterministic}. It is important to realize that functions $W_t^*$ and $V_t^*$ can be anything at all that insurers find useful and practical, and regulators allow or require. For example, an older practice was to take assets at book values rather than market values. More recently, perfect matching or a self-financing condition would require that $W_t^*({\cal H}_t) = V_t^*({\cal H}_t)$ at all times $t$; equation (\ref{eq:EqPrinPrem1.5}) but in a stochastic world. A first step towards achieving mathematical coherence in practice is to find a framework in which practices can be expressed.

%--------------------------------------------------------------

\subsection{The Stochastic Time-evolution of the Balance Sheet}
\label{sec:Stochastic Evolution}

Panel (b) of Figure \ref{fig:TechBases} illustrates the position half-way through the policy term. Asset values are accumulations of  past cashflows; two possible sample paths or scenarios are shown\footnote{The stochastic scenarios shown in Figure \ref{fig:TechBases} are merely indicative of stochastic irregularity, based on a large portfolio of policies, they have not been generated by any of the models discussed here.}. Liabilities are the in-force policy values, shown on three different technical bases as before. The difference between assets and liabilities at $t=10$ is the accumulated surplus. Panel (c) shows final outcomes at $t=20$ in both stochastic scenarios.

%-------------------------------------------------------------
%-------------------------------------------------------------
%-------------------------------------------------------------

\section{Notation: Life History, Cashflow, Interest, Probability, {Information}}
\label{sec:NotationsRepresentations}

\subsection{Representing Life Histories}
\label{sec:LifeHistories}

There are three ways to represent the future lifetime of ($x$), assuming death to be the only stochastic event. We use whichever is convenient, and call it the biometric model.

\begin{bajlist} 

\item {\em Random Lifetime $T$}: The time until death is represented by a non-negative random variable $T$. 

\item {\em Counting Process $N(t)$}: The life history is represented by the right-continuous counting process $N(t)$ defined as:

\begin{equation}
N(t) = 1_{\{ T \le t \}} = \left\{ \begin{array}{ll}0 & \mbox{if } T > t \\ 1 & \mbox{if } T \le t. \end{array} \right. \label{eq:DefN}
\end{equation}

\item {\em State Process $J(t)$}: Define the state space $\mathcal{S} = \{1,2\}$ representing `alive' and `dead' statuses  and a right-continuous process $J(t)$ as follows:  

\begin{equation}
J(t) = \mbox{state in ${\cal S}$ occupied at time $t$} \qquad (t \ge 0). \label{eq:StateProcessDef}
\end{equation}

\noindent Then suppose that $J(t)$ jumps from state 1 to state 2 at time $T$. This extends in an obvious way to a multiple-state model on a larger state space $\mathcal{S} = \{ 1, 2, \ldots, K \}$. 

\end{bajlist}

\noindent {Associated with the event of death is the right-continuous indicator of being alive at time $t$ which is defined in terms of any of the above representations\footnote{In this paper we use $1_{\{ N(t)=0 \}}$ simply because so many expressions are written in terms of $dN(t)$. In multiple-state models the form $1_{\{ J(t)=1 \}}$ generalizes most easily.} as:

\begin{equation}
1_{\{ T > t \}} = 1_{\{ N(t)=0 \}} = 1_{\{ J(t)=1 \}}. \label{sec:Indicators}
\end{equation}

\noindent In survival analysis the left-continuous $1_{\{ T \ge t \}} = 1_{\{ N(t^-)=0 \}} = 1_{\{ J(t^-)=1 \}}$ is used instead because of its r\^ole in estimators written as stochastic integrals.} We have the identities:

\begin{equation}
dN(t) = - d\{ 1_{\{ N(t)=0 \}} \} = \big( 1_{\{ T \ge t \}} - 1_{\{ T > t \}} \big). \label{eq:CPIdentities}
\end{equation}

%----------------------------------------------------------------

\subsection{Representing Contractual Cashflows}
\label{sec:ContractualCashflows}

We assume a life insurance contract commences at time $t=0$ and has finite term $n$ years. Let $B(t)$ $(0 \le t \le n)$ be the stochastic cumulative cashflow generated on the interval $[0,t]$ by the contractual payments, with the convention that cashflows to the insurer are positive. Randomness in $B(t)$ derives from the biometric model in Section \ref{sec:LifeHistories}, and lump-sum payments at non-random times are denoted by $\Delta B(t) = B(t) - B(t^-)$. For simplicity we ignore such payments except at $t=n$, in particular we suppose that $B(0)=0$. We assume that $B(t)$ may represented by the Stieltjes integral:

\begin{equation}
B(t) = \int_0^t dB(r) \qquad (0 \le t \le n). \label{eq:NorbergStage1}
\end{equation}

\noindent To make the presentation even simpler we define a rate of premium $P_t$ and death benefit $S_t$ for $t \in (0,n)$, and maturity benefit $\bar{S}$ at $t=n$. Therefore: 

\begin{equation}
dB(t) = \left\{ \begin{array}{ll} 1_{ \{ N(t)=0 \}} \, P_t \, dt - S_t \, dN(t) & \quad t < n \\[0.5ex] 1_{ \{ N(n)=0 \}} \, P_t \, dt - S_n \, dN(n) -1_{\{ N(n)=0 \}} \, \bar{S} & \quad t=n. \end{array} \right. \label{eq:CPRep}
\end{equation}

Traditional actuarial value functions are defined `per-policy now in force'. The indicators in equation (\ref{eq:CPRep}) means cashflows are `per policy in force at $t=0$'.

%----------------------------------------------------------------

\subsection{Representing Interest, Accumulation and Discounting}
\label{sec:RepresentingInterest}

We assume a left-continuous deterministic {force of interest} $\delta_t$, such that $\exp \big(\int_0^t \delta_r \, dr \big)$ represents the amount to which \$1 invested at time 0 accumulates at time $t$. The {discount factor} $v(t)$ is defined as $\exp \big(-\int_0^t \delta_r \, dr \big)$. We comment on stochastic asset returns in Section \ref{sec:FinancialInfo}.

%----------------------------------------------------------------

\subsection{Distribution Function and Hazard Rate}
\label{sec:RepresentingProbs}

The distribution function of $T$, denoted by $F(t)$, is:

\begin{equation}
F(t) = \Prob[T \le t]. \label{eq:DistDef}
\end{equation}

\noindent The {hazard rate} (alternatively {intensity, force of mortality}) denoted by $\mu_t$ is then:

\begin{equation}
\mu_t = \lim_{dt \to 0} \frac{\Prob[N(t+dt) = 1 \mid N(t)=0]}{dt} = \lim_{dt \to 0} \frac{1}{dt} \, \frac{F(t+dt) - F(t)}{1 - F(t)} \label{eq:DefMu}
\end{equation}
 
\noindent which is assumed always to exist\footnote{\label{foot:Bounded}Boundedness and integrability of transition intensities on closed intervals is enough to rule out `explosions' in models which can jump more than once \citep{jacobsen2006}. Here, it allows the hazard to be unbounded as $t \to \infty$ (for example, a Gompertz-Makeham model) but rules out models with a finite limiting lifetime ($F(t) = 1$ for all $t > \omega$ with $\omega < \infty$). This restriction is easily removed.}. 

%----------------------------------------------------------------

\newpage

\subsection{Representing Information}
\label{sec:RepresentingInfo}

\subsubsection{Biometric Risk}
\label{sec:BiometricInfo}

We assume a filtered probability space ${\cal S} = ( \Omega, {\cal F}, \mathbf{F} , \mathbb{P} )$, where $\mathbf{F} =  \{ \mathcal{F}_t \}_{ t \ge 0 }$ is a filtration with ${\cal F} = {\cal F}_{\infty} = \lim_{t \to \infty} {\cal F}_t$. We suppose the points $\omega \in \Omega$ are sample paths of $N(t)$, also called {scenarios}, and list some {assumptions} below.

\begin{bajlist}

\item The filtration is the natural filtration generated by $N(t)$, ${\cal F}_t = \sigma \big( \{ N(s) \}_{0 \le s \le t} \big)$, so $N(t)$ {and $1_{\{ N(t)=0 \}}$ are ${\cal F}_t$-measureable}. 

\item The measure $\mathbb{P}$ is {that induced by} the distribution function $F(t)$ or hazard rates $\mu_t$. 

\item The Markov property is conferred on the process and its filtration (or not) by $\mathbb{P}$, and is satisfied if:

\begin{equation}
\Mean_{\mathbb{P}}[ \, A \mid {\cal F}_t \, ] = \Mean_{\mathbb{P}}[ \, A \mid N(t) \, ] \qquad \mbox{for all } A \in {\cal F}. \label{eq:MarkovProperty}
\end{equation}

\item The {martingale} property holds:

\begin{equation}
\Mean_{\mathbb{P}}[ \, dN(t) \mid {\cal F}_{t} \, ] = \Mean_{\mathbb{P}}[ \, 1_{\{ N(t)=0 \}} \, \mu_t \, dt \mid {\cal F}_{t} \, ] = 1_{\{  N(t)=0 \}} \, \mu_t \, dt \label{eq:MartingalePropA}
\end{equation}

\noindent equivalently, the process $M(t)$ defined as $N(t) - \int_0^t 1_{\{ N(r)=0 \}} \, \mu_r \, dr$ is an ${\cal F}_t$-martingale under $\mathbb{P}$. 

\end{bajlist}

%---------------------------------------------------------------

\subsubsection{Financial Risk}
\label{sec:FinancialInfo}

Ideally, the filtration $\mathbf{F}$ would be subsidiary to a larger filtration $\mathbf{H}$ in a stochastic model of financial and biometric risk. If we had a stochastic asset model with natural filtration $\mathbf{G}$, we might combine the underlying spaces in such a way that they support a combined filtration $\mathbf{H} = \mathbf{F} \vee \mathbf{G}$, and define stochastic models by probability measures on ${\cal H}_{\infty}$. This approach has indeed been explored, see for example \cite{asmussen2020} but little progess has been made beyond assuming  financial and biometric risk to be independent\footnote{\cite{norberg1999} provided a {stochastic environment} by allowing the parameters of financial and biometric models (here, ($\delta_t, \mu_t$)) to be a stochastic process, generating the  filtration $\mathbf{G} =  \{ \mathcal{G}_t \}_{ t \ge0 }$.  Norberg (p.379) assumed that: ``$\ldots$ the sample paths of the processes $\delta$ and $\mu_{jk}$ almost surely possess the properties assumed in Section 2 $\ldots$'' (namely a Markov model in which the Kolmogorov equations hold, and interest that may be expressed through a discount factor $\exp ( -\int \delta_r \, dr )$). Then all expectations at time $t$ were conditioned on $\mathcal{F}'_t \vee \mathcal{G}_t$ where $\mathcal{F}'_t \subseteq \mathcal{F}_t$. The implication of conditioning on $\mathcal{G}_t$ was that the realized $(\delta_r , \mu_r)$ for $r \in [0,t]$ could be treated as if it was deterministic; in other words, carry on as before. Norberg pointed out that ``$\ldots$ no particular specification of the marginal distribution [of $(\delta_r , \mu_r)$] is needed $\ldots$''.}. In the text we assume $\mathbf{G}$ is trivial. Like multiple states, a stochastic asset model would lead to {more complex expressions, but the gain in insight would be in a different direction;} by omitting it we do not mean to underrate its importance.

%-------------------------------------------------------------
%-------------------------------------------------------------
%-------------------------------------------------------------
%-------------------------------------------------------------

\section{Technical Bases and Thiele's Differential Equation}
\label{sec:TechBasesOverall}
 
\subsection{Defining Technical Bases}
\label{sec:TechBases}

We follow earlier authors by defining a generic technical basis denoted by ${\cal B}^Z$, to be the pair $(\delta_t^Z,\mu_t^Z)$ of (vector) parameters specifying, respectively, investment and biometric models. The index $Z$ labels different technical bases ${\cal B}^X, {\cal B}^Y$, and so on. 

\begin{bajlist}

\item Any technical basis ${\cal B}^Z$ is assumed to represent (induce) a probability (measure) $\Prob_Z$ {equivalent to the reference measure $\mathbb{P}$ {induced as in Section \ref{sec:BiometricInfo}(b)}, also a distribution $F_Z(t)$ of $T$ (equation (\ref{eq:DistDef}));} and an expectation operator $\Mean_Z$.

\item Given cashflows $S_t$ and $P_t$, technical basis ${\cal B}^Z$ parametrizes Thiele's differential equation \citep{hoem1988} as follows:

\begin{equation}
\frac{d}{dt} \, g(t) = \delta_t^Z \, g(t) + P_t - \mu_t^Z \, ( S_t - g(t) ). \label{eq:ThieleGeneric}
\end{equation}

\item The maturity benefit $\bar{S}$ does not appear in equation (\ref{eq:ThieleGeneric}); its r\^ole is as a boundary condition of certain solutions, see Section \ref{sec:ThieleSolutions}.

\item The quantity $S_t - g(t)$ appearing in Thiele's equation is called the sum at risk.  

\item Under ${\cal B}^Z$, the process $M^Z(t) \equiv N(t) - \int_0^t 1_{\{ N(r) = 0 \}} \, \mu_r^Z \, dr$ is an ${\cal F}_t$-martingale.

\end{bajlist}

%-------------------------------------------------------------

\subsection{Solutions of Thiele's Equation: Policy Values, Accumulations and Sums at Risk}
\label{sec:ThieleSolutions}

It is standard to check the following solutions of Thiele's equation (\ref{eq:ThieleGeneric}), conveniently expressed in terms of the following discounting/survival factors, see equation (\ref{eq:IntegratingFactor}):

\begin{equation}
v^Z(t) = \exp \left( - \int_0^t \delta_r^Z \, dr   \right), \quad p^Z(t) = \exp \left( - \int_0^t \mu_r^Z \, dr   \right), \quad \varphi^Z(t) = v^Z(t) \, p^Z(t). \label{eq:IntegratingFactorZ}
\end{equation}

\begin{bajlist}

\item {\em Backward solution}: Solving (\ref{eq:ThieleGeneric}) backwards from the boundary value $g(n)=\bar{S}$ gives:

\begin{eqnarray}
% g(t) & = & \int_t^n e^{-\int_t^r (\delta^Z_s + \mu^Z_s) ds} \, (\mu_r^Z \, S_r - P_r) \, dr + e^{-\int_t^n (\delta^Z_s + \mu^Z_s) ds} \, \bar{S} \nonumber \\
g(t) & = & \int_t^n \frac{\varphi^Z(r)}{\varphi^Z(t)} \, (\mu_r^Z \, S_r - P_r) \, dr + \frac{\varphi^Z(n)}{\varphi^Z(t)}  \, \bar{S}. \label{eq:ThieleSolBackwards}  
\end{eqnarray}

\noindent In this case we usually denote the technical basis by ${\cal B}^L$ and call it a valuation basis; denote the solution by $V_t^L$ and call it a policy value; and denote the associated sum at risk by $R^L(t) = S_t - V_t^L$. % See Section \ref{sec:InterpretationPolVals} for the {r\^ole of $V_t^L$ in defining a conditional expectation}. 

\item {\em Forward solution}: Solving (\ref{eq:ThieleGeneric}) forwards from the boundary value $g(0)=0$ gives:

\begin{eqnarray}
% g(t) & = & \int_0^t e^{\int_r^t (\delta^Z_s + \mu^Z_s) ds} \, (P_r - \mu_r^Z \, S_r) \, dr \nonumber \\
g(t) & = & \int_0^t \frac{\varphi^Z(r)}{\varphi^Z(t)}  \, (P_r - \mu_r^Z \, S_r) \, dr. \label{eq:ThieleSolForwardsA} 
\end{eqnarray}

\noindent In this case we usually denote {the technical basis by ${\cal B}^A$,} and call it an accumulation basis; denote the solution by $W_t^A$ and call it an accumulation\footnote{Traditionally, $V_t^L$ and $W_t^A$ would be called  the prospective and retrospective policy values, respectively; see the Appendix of \cite{hacariz2024} for a discussion of the latter terminology. We follow \cite{linnemann2002, linnemann2003} in adopting the `$W$' notation for accumulations.}; and denote the associated {sum at risk} by $R^A(t) = S_t - W_t^A$. 

\item A solution of equation (\ref{eq:ThieleGeneric}) satisfying both $g(0)=0$ and $g(n)= \bar{S}$ satisfies the principle of equivalence; we call the technical basis a special valuation basis. For a given technical basis there may be infinitely many premium rate functions $P_t$ satisfying the equivalence principle. In practice we assume there are constraints that identify a unique solution, for example, a level premium rate payable throughout the term.

\item The (partial) classification of technical bases provided by (a) to (c) above is summarized in Table \ref{table:TechnicalBases}, following \cite{hacariz2024}.

\end{bajlist}

%-------------------------------------------------------------

\begin{table}
\begin{center}
\caption{\label{table:TechnicalBases} Characteristics of technical bases in terms of boundary conditions satisfied by  solutions of Thiele's differential equation. Solutions are deterministic functions of time $t$.}
\small
\begin{tabular}{lllll}
& & & & \\
%  &      &          &          &  \\
{Basis} & Name & Type & {Solution} & Boundary Condition(s)  \\[0.5ex]
${\cal B}^P$ & Premium & Special Valuation & $V_t^P$ & $V_0^P = 0, V_n^P = \bar{S}$ \\
${\cal B}^L$ & Valuation & Valuation & $V_t^L$ & $V_n^L = \bar{S}$ \\
${\cal B}^A$ & Accumulation & Accumulation & $W_t^A$ & $W_0^A = 0$ \\
${\cal B}^M$ & Experience & Accumulation & $W_t^M$ & See Section \ref{sec:TechBasisBM}.
\end{tabular}
\end{center}
\end{table}

%-------------------------------------------------------------

\subsection{The Experience Technical Basis ${\cal B}^M$}
\label{sec:TechBasisBM}

We suppose a `true' technical basis exists, which generates the stochastic cashflows $B(t)$, and denote it by  ${\cal B}^M$ (`M' stands for `martingale'). We reserve the term `experience basis' exclusively to ${\cal B}^M$, and use ${\cal B}^A$ to denote any other {accumulation basis}. When we model the actuary's actions from a higher standpoint (for example in a model office) we may treat ${\cal B}^M$ as a meta-model which we know but the model actuary does not. 

%-------------------------------------------------------------
%-------------------------------------------------------------
%-------------------------------------------------------------

\section{The Balance Sheet}
\label{sec:BalanceSheet}

\subsection{A Stochastic Model Balance Sheet}
\label{sec:AssetsLiabilities}

The balance sheet displays the values placed on the assets and the liabilities, and the difference between them is surplus. Barring exact cashflow matching, which would reduce the balance sheet to a tautology, surplus depends on valuation methodology. 

In this section and the next, we assume that there is only one technical basis, so without ambiguity we can drop postscripts denoting technical bases and refer to technical basis ${\cal B} = (\delta_t,\mu_t)$, discount functions $v(t)$, $p(t)$ and $\varphi(t)$, conditional expectations $\Mean[ \, \bullet \mid {\cal F}_t \, ]$, martingale $M(t)$, policy values $V_t$ and accumulations $W_t$.

Follow \cite{norberg1991} and define ${\cal F}_n$-measureable $X(t)$ as the value at time $t$ of all cashflows, recalling that $B(0)=0$, and split $X(t)$ into past and future parts:

\begin{eqnarray}
X(t) & = & {\int_{0}^n \frac{v(r)}{v(t)} \, dB(r)} \label{eq:NorbergCashflows} \\
& = & \underbrace{\int_{0}^t \frac{v(r)}{v(t)} \, dB(r)}_{\mbox{\scriptsize Accum. Past Cashflows}} - \underbrace{\int_{t}^n \frac{v(r)}{v(t)} \, d(-B)(r)}_{\mbox{\scriptsize Disc. Future Cashflows}}  \label{eq:NorbergStage2} 
\end{eqnarray}

\noindent where, as usual, we write $\int_s^t$ to mean $\int_{(s,t]}$ to save space. If we then follow \cite{buehlmann1976} by defining the propective policy value at time $t$ to be the conditional expectation:

\begin{equation}
\Mean \left[ \int_t^n \frac{v(r)}{v(t)} \, d(-B(r)) \, \Big| \, {\cal F}_t \right] \label{eq:LiabilityValueL}
\end{equation}

\noindent we get a simple stochastic model of the balance sheet:

\begin{eqnarray}
\Mean[ \, X(t) \mid {\cal F}_t \, ] & = & \Mean \left[ \int_0^t \frac{v(r)}{v(t)} \, dB(r) \, \Big| \, {\cal F}_t \, \right] - \Mean \left[ \int_t^n \frac{v(r)}{v(t)} \, d(-B(r)) \, \Big| \, {\cal F}_t \right]  \label{eq:ModelBalSht1} \\[1.0mm]
& = & \underbrace{\int_0^t \frac{v(r)}{v(t)} \, dB(r)}_{\mbox{\scriptsize Asset Value}} - \underbrace{\Mean \left[ \int_t^n \frac{v(r)}{v(t)} \, d(-B(r)) \, \Big| \, {\cal F}_t \right]}_{\mbox{\scriptsize Prospective Policy Value}}. \label{eq:ModelBalSht2}
\end{eqnarray}

However, we now have two quantities named `prospective policy value'; one a deterministic function $V_t$ obtained as a solution of Thiele's equation, the other the ${\cal F}_t$-conditional expectation in equation (\ref{eq:LiabilityValueL}). Proposition \ref{prop:IdentityV} below shows how they are related.

%-------------------------------------------------------------

\subsection{{Stochastic} Interpretation of Policy Values {and Accumulations}}
\label{sec:InterpretationPolVals}

\begin{proposition} \label{prop:IdentityV} If $V_t$ is a solution of Thiele's equation (\ref{eq:ThieleGeneric}) satisfying  $V_n=\bar{S}$, then:

\begin{equation}
1_{\{ N(t)=0 \}} \, V_t = \Mean \left[ \int_t^n \frac{v(r)}{v(t)} \, d(-B(r)) \, \Big| \, {\cal F}_t \right]. \label{eq:ProspPolValasCondExp}
\end{equation}

\end{proposition}

\noindent Both sides of this equation are ${\cal F}_t$-measureable random variables. The proposition says that a stochastic policy value, produced by multiplying deterministic $V_t$ by stochastic $1_{\{ N(t)=0 \}}$, is expressible as an ${\cal F}_t$-conditional expectation of discounted future cashflows; {equivalently takes constant values on the sets $\{N(t)=0\}$ and $\{N(t) =1\}$}. 

\medskip

\noindent {\em Proof}: Consider $d \big\{ v(t) \, 1_{\{ N(t)=0 \}} \, V_t \big\}$;

\begin{eqnarray}
d \big\{ v(t) \, 1_{\{ N(t)=0 \}} \, V_t \big\} & = & -\delta_t \, v(t) \, 1_{\{ N(t)=0 \}} \, V_t \, dt + v(t) \, d \big\{ 1_{\{ N(t)=0 \}} \, V_t \big\} \nonumber \\
& = & -\delta_t \, v(t) \, 1_{\{ N(t)=0 \}} \, V_t \, dt + v(t) \, d \big\{ 1_{\{ N(t)=0 \}} \big\} \, V_t \nonumber \\ 
&   & \quad + v(t) \, 1_{\{ N(t)=0 \}} \, \big\{ \delta_t \, V_t + P_t - \mu_t \, ( S_t - V_t ) \big\} \, dt. \label{eq:ProspPolValasCondExpProofA}
\end{eqnarray}

\noindent Applying the first identity in equation (\ref{eq:CPIdentities}) at the middle term above, then adding and subtracting $v(t) \, S_t \, dN(t)$ we have:

\begin{eqnarray}
d \big\{ v(t) \, 1_{\{ N(t)=0 \}} \, V_t \big\} & = & v(t) \, \big\{ 1_{\{ N(t)=0 \}} \, P_t \, dt - S_t \, dN(t) \big\} \nonumber \\
&   & + v(t) \, (S_t - V_t) \, \big\{ dN(t) - 1_{\{ N(t)=0 \}} \, \mu_t \, dt \big\}. \label{eq:ProspPolValasCondExpProofC}
\end{eqnarray}

\noindent Integrate over $r \in [t,n]$, and divide by $v(t)$:

\begin{eqnarray}
\frac{v(n)}{v(t)} \, 1_{\{ N(n)=0 \}} \, V_n - \frac{v(t)}{v(t)} \, 1_{\{ N(t)=0 \}} \, V_t & = &  \int_t^n \frac{v(r)}{v(t)} \, \big( 1_{\{ N(r)=0 \}} \, P_r \, dr - S_r \, dN(r) \big) \nonumber \\
&   & \quad + \int_t^n \frac{v(r)}{v(t)} \, (S_r - V_r) \, dM(r).  \label{eq:ProspPolValasCondExpProofF}
\end{eqnarray}

\noindent Note that $V_n = \bar{S}$, rearrange, and take conditional expectations $\Mean[ \, \bullet \mid \, {\cal F}_t \, ]$. 
\hfill{$\Box$}

\smallskip

\noindent {The converse of Proposition \ref{prop:IdentityV}, that Thiele's equation is satisfied by: 

\begin{equation}
g(t) = \Mean \left[ \int_t^n \frac{v(r)}{v(t)} \, d(-B(r)) \, \Big| \, {\cal F}_t , N(t)=0 \right] \label{eq:ProspPolValasStatewiseCondExp}
\end{equation}

\noindent is easily proved by direct calculation (see \cite{norberg1991}). In the retrospective case we have the following result.}

\begin{corollary} \label{corr:IdentityA} If $W_t$ is a solution of Thiele's equation satisfying $W_0=0$, then:

\begin{equation}
W_t = \frac{1}{\varphi(t)} \, \Mean \left[ \int_0^t v(r) \, dB(r) \, \Big| \, {\cal F}_0 \right]. \label{eq:ThieleSolForwards}
\end{equation}

\end{corollary}

\noindent Both sides of this equation are ${\cal F}_0$-measureable random variables. 

\smallskip

\noindent {\em Proof}: Consider $d \big\{ v(t) \, 1_{\{ N(t)=0 \}} \, W_t \big\}$, and follow the same steps as far as equation (\ref{eq:ProspPolValasCondExpProofC}). Then integrate over $r \in [0,t]$, divide by $v(t)$, note that $W_0=0$ and take conditional expectations $\Mean[ \, \bullet \mid \, {\cal F}_0 \, ]$. \hfill{$\Box$}

\smallskip

The `scale' of both $V_t$ and $W_t$ is `per policy in-force at time $t$', which explains the presence of indicator $1_{\{ N(t)=0 \}}$ in equation (\ref{eq:ProspPolValasCondExp}), and factor $\varphi(t)^{-1}$ in equation (\ref{eq:ThieleSolForwards}).

%-------------------------------------------------------------

\subsection{Asset and Liability Valuation Functions Again}
\label{sec:AssetsLiabilitiesII}

We can now interpret the stochastic balance sheet model of equation (\ref{eq:ModelBalSht2}):

\begin{eqnarray}
\Mean[ \, X(t) \mid {\cal F}_t \, ] & = & \underbrace{\int_0^t \frac{v(r)}{v(t)} \, dB(r)}_{\mbox{\scriptsize Asset Value}} - \underbrace{1_{\{ N(t)=0 \}} \, V_t}_{\mbox{\scriptsize Prosp. Pol. Val.}}. \label{eq:ModelBalSht3} 
\end{eqnarray}

\noindent The prospective policy value is an ${\cal F}_t$-measureable random variable, {taking the deterministic value $V_t$ if the insured individual is alive, and the deterministic value 0 if not.} 

%-------------------------------------------------------------

\cite[Section 1A]{norberg1991} said: ``The concept of prospective reserve is no matter of dispute in life insurance mathematics. It is defined as the conditional expected present value of future benefits less premiums, given its present state.'' This is consistent with \cite{buehlmann1976} and (as we will see) with Scandinavian-style regulation, but it is too limiting, for example it excludes all the approaches used after the Northampton method was abandoned by UK actuaries. {Rather, we can now flesh out the asset and liability valuation functions mentioned in Section \ref{sec:ValuationFunctions}. If we are to work with them mathematically we hope that {each is a function of a technical basis, a stochastic cashflow function and a flow of information, chosen for the particular purpose}. Omitting formal definitions, they might look like this:}

\begin{eqnarray}
\mbox{Asset Value} & = & W_t^* \big( {\cal B}^A, \{ dB^A(t) \}_{0 \le t \le n}, {\cal F}^A_t \big)  \label{eq:AssetsGeneral} \\
\mbox{Liability Value} & = & V_t^* \big( {\cal B}^L, \{ dB^L(t) \}_{0 \le t \le n}, {\cal F}^L_t \big) . \label{eq:LiabilityGeneral} 
\end{eqnarray}

\noindent By choosing the two terms on the right-hand side of equation (\ref{eq:ModelBalSht2}) for the functions $W_t^*$ and $V_t^*$, {we choose} Scandinavian-style regulation. 

%-------------------------------------------------------------
%-------------------------------------------------------------
%-------------------------------------------------------------

\section{One Technical Basis: Introducing Surplus}
\label{sec:modelingBalanceSheet}

\subsection{Plan of Approach}
\label{sec:PlanApproach}

The definition and analysis of surplus proceeds in two parts. 

\begin{bajlist}

\item {\em One technical basis}. With one technical basis ${\cal B}$ the surplus at time $t$ is given by equation (\ref{eq:ModelBalSht3}). Further discounting to $t=0$ and/or taking ${\cal F}_0$-conditional expectations of the surplus shows what projected models of future surplus may be available to the actuary at the inception of the contract. We set out these steps in this section, and in addition define a useful notation for surplus. 

\item {\em Two and three technical bases}.

\begin{bajsublist}

\item In Section \ref{sec:StochAndmodeled}, we model Scandinavian-style regulation with two technical bases, a `first-order' ${\cal B}^L$ common to premium and policy value calculations, and a `second-order' ${\cal B}^M$ describing the  experience (see Section \ref{sec:TechBasisBM}). Surplus has a systematic or planned component, resulting from a safe-side first-order technical basis.

\item In Section \ref{sec:DiffPremValBases} we allow for separate premium, valuation and experience technical bases ${\cal B}^P, {\cal B}^L$ and ${\cal B}^M$. Each pair of technical bases defines a different property of the surplus, as described in \cite{hacariz2024}. However some valuation methods (including the historically important net premium method) go beyond choosing a technical basis (meaning that the technical basis ${\cal B}^L$ is not the only component of the valuation function $V_t^*$ in equation (\ref{eq:LiabilityGeneral})). We define an actuarial basis as an enlarged technical basis in Section \ref{sec:TechBasisCashflows} and the resulting surplus in Section \ref{sec:ModifiedDecomp}.

\end{bajsublist}

\end{bajlist}

%------------------------------------------------------------------

\subsection{Stochastic and Modeled Surplus with One Technical Basis}
\label{sec:Norberg}

We suppose the technical basis is a special valuation basis so $V_0=0$ and $V_n = \bar{S}$.

%------------------------------------------------------------------

\subsubsection{Stochastic Surplus}
\label{sec:StochSurp}

The conditional expectation in equation (\ref{eq:ModelBalSht3}) is the stochastic surplus, undiscounted: 

\begin{eqnarray}
\Mean[X(t) \mid {\cal F}_t] & = & \Mean \left[ \int_0^t \frac{v(r)}{v(t)} \, dB(r) \, \Big| \, {\cal F}_t \right] -  \Mean \left[ \int_t^n \frac{v(r)}{v(t)} \, d(-B)(r) \, \Big| \, {\cal F}_t \right] \label{eq:NorbergStage3a} \\[0.5ex]
& = & \int_0^t \frac{v(r)}{v(t)} \, dB(r) - 1_{\{ N(t)=0 \}} \, V_t \label{eq:NorbergStage3b}
\end{eqnarray}

\noindent and it is sometimes useful to discount this surplus to $t=0$:

\begin{equation}
v(t) \, \Mean[ X(t) \mid {\cal F}_t] = \int_0^t {v(r)} \, dB(r) - v(t) \, 1_{\{ N(t)=0 \}} \, V_t.\label{eq:NorbergStage3c}
\end{equation}

\noindent These are ${\cal F}_t$-measurable random variables. 

%------------------------------------------------------------------

\subsubsection{Modeled Surplus}
\label{sec:ModSurp}

The conditional ${\cal F}_0$-expectations of equations (\ref{eq:NorbergStage3b}) and (\ref{eq:NorbergStage3c}) see \cite[equation (5.6)]{norberg1999} we call the {modeled surplus}. It may be defined undiscounted: 

\begin{eqnarray}
\Mean[ \, \Mean[X(t) \mid {\cal F}_t] \mid {\cal F}_0] 
& = & \Mean \left[ \int_0^t \frac{v(r)}{v(t)} \, dB(r) \, \Big| \, {\cal F}_0 \right] - \Mean[ 1_{\{ N(t)=0 \}} \, V_t \mid {\cal F}_0] \label{eq:NorbergStage4a} \\[1.0ex]
& = & p(t) \, ( W_t - V_t ) \label{eq:NorbergStage4aa}
\end{eqnarray}

\noindent or with financial discounting:

\begin{eqnarray}
\Mean[ v(t) \, \Mean[X(t) \mid {\cal F}_t] \mid {\cal F}_0] 
& = & \Mean \left[ \int_0^t {v(r)} \, dB(r) \, \Big| \, {\cal F}_0 \right] - \Mean[ v(t) \, 1_{\{ N(t)=0 \}} \, V_t \mid {\cal F}_0]  \label{eq:NorbergStage4} \\[1.0ex]
& = & {\varphi(t) \, ( W_t -  V_t ).} \label{eq:NorbergStage4ab}
\end{eqnarray}

\noindent These are ${\cal F}_0$-measurable random variables. In particular $W_t$, which is the classical retrospective reserve, is an ${\cal F}_0$-conditional expectations, despite its name. The fact that equations (\ref{eq:NorbergStage4aa}) and (\ref{eq:NorbergStage4ab}) are identically zero, which we show in Corollary \ref{corr:IntegrateSurp1}, expresses Norberg's well-known remark that the classical retrospective reserve is: ``$\ldots$ rather a retrospective formula for the prospective reserve'' \citep{norberg1991}.

\medskip

%------------------------------------------------------------------

\subsubsection{The Function $\Mean[ \, X(t) \mid {\cal F}_t \, ]$}
\label{sec:FunctionComments}

The object of interest is $\Mean[ \, X(t) \mid {\cal F}_t \, ]$ as a function of $t$. We never consider $\Mean[ \, X(t) \mid {\cal F}_r \, ]$ for $r \not= t$. Note the following.

\begin{bajlist}

\item Derivatives of $\Mean[ \, X(t) \mid {\cal F}_t \, ]$ will involve $dX(t)$, where $X(t)$ has the form $\int_0^t - \int_t^n$. At this point the two integrands have the same value at their respective upper and lower limits of integration (time $t$) see equation (\ref{eq:NorbergStage2}). This will not always be so.

\item Although the left-hand side of equation (\ref{eq:NorbergStage4a}) simplifies as $\Mean[ \, X(t)\mid {\cal F}_0 \, ]$, this is of interest only for $t=0$, where $\Mean[ \, X(0)\mid {\cal F}_0 \, ] = 0$  expresses the equivalence principle. 

\end{bajlist}

%------------------------------------------------------------------

\subsubsection{Comments}
\label{sec:SurpComments}

\begin{bajlist}

\item Figure \ref{fig:TechBases} illustrates the idea of learning past history and needing to model only the future. Panel (a) shows ${\cal F}_0$-conditional expectations (Section \ref{sec:ModSurp}) when there is no history. Panel (b) shows ${\cal F}_t$-conditional expectations (Section \ref{sec:StochSurp}) as time passes. At time $n$ everything is known; panel (c) shows entire ${\cal F}_n$-measurable sample paths. 

\item The given definitions of surplus do not imply that $V_0=0$ or $V_n=\bar{S}$. For example, they do not  imply that the policy values in panel (a) of Figure \ref{fig:TechBases} should start and end at zero. It is a matter of {assumption} in this section that these conditions all hold. 

\item If the cashflows $B(t)$ represented a self-financing financial portfolio with $X(n)=0$, not only the ${\cal F}_0$-measurable quantities in Section \ref{sec:ModSurp}, but all the ${\cal F}_t$-measurable quantities in Section \ref{sec:StochSurp} would be identically zero. In other words, the balance sheet would record no more than hedging error. 

\end{bajlist}

%----------------------------------------------------------------
%----------------------------------------------------------------

\begin{table}
\begin{center}
\caption{\label{table:ThetaNotation} `$\Theta$' notation for stochastic and modeled surplus equations at time $t$, with a single (generic) technical basis ${\cal B}^Z$ and value of contractual cashflows $X(t)$.}
\small
\begin{tabular}{lllll}
& & & & \\
Name & Notation & Definition & Meas. & Equation \\[1.0ex]
Realized   & $\Theta^{\omega,\omega}(t)$ & $X(t)$ & ${\cal F}_n$ & (\ref{eq:NorbergStage2}) \\[0.5ex]
Stochastic & $\Theta^{Z,\omega}(t)$ & $\Mean_Z[ \, X(t) \mid {\cal F}_t \, ]$ & ${\cal F}_t$ & (\ref{eq:NorbergStage3a}),  (\ref{eq:NorbergStage3b}) \\[0.5ex]
Stochastic discounted & $\tilde{\Theta}^{Z,\omega}(t)$ & $v^Z(t) \, \Mean_Z[ \, X(t) \mid {\cal F}_t \, ]$ & ${\cal F}_t$ & (\ref{eq:NorbergStage3c}) \\[0.5ex]
Modeled & ${\Theta}^{Z,Z}(t)$ & $\Mean_Z[ \, \Mean_Z[ \, X(t) \mid {\cal F}_t \, ] \mid {\cal F}_0 \, ]$ & ${\cal F}_0$ & (\ref{eq:NorbergStage4a}), (\ref{eq:NorbergStage4aa}) \\[0.5ex]
Modeled discounted & $\tilde{\Theta}^{Z,Z}(t)$ & $\Mean_Z[ \, v^Z(t) \, \Mean_Z[ \, X(t) \mid {\cal F}_t \, ] \mid {\cal F}_0 \, ]$ & ${\cal F}_0$ & (\ref{eq:NorbergStage4}), (\ref{eq:NorbergStage4ab}).
\end{tabular}
\end{center}
\end{table}

%----------------------------------------------------------------

\subsection{Notation for Stochastic and Modeled Surplus}
\label{sec:NotationBalShts}

We introduce {`$\Theta^{Z_1,Z_2}(t)$'} notation\footnote{This notation extends the `$\Gamma$' notation from \cite{ramlau-hansen1988a} and \cite{hacariz2024}. The change from `$\Gamma$' to `$\Theta$' is for purely typographical reasons, it is easier on the eye with superscripts added.} to represent surplus in models following \cite{buehlmann1976} and \cite{norberg1991} in choosing conditional expectations as the valuation function.   

\begin{bajlist}

\item The first superscript place shows the basis used to discount and condition over $(t,n]$.

\item The second superscript place shows the basis of discount and condition over $[0,t]$.

\item Symbol `$\omega$' in the superscript indicates the actual sample path (no conditioning). 

\item A tilde `$\sim$' indicates discounting to $t=0$. 

\item {We include the value of realized surplus $\Theta^{\omega,\omega}(t) = X(t)$ for completeness only.}

\end{bajlist}

\noindent The differentials $d \Theta^{Z_1,Z_2}(t)$ or $d \tilde{\Theta}^{Z_1,Z_2}(t)$ model the rate of surplus generation at time $t$, being shorthand respectively for a stochastic differential equation and an ordinary differential equation, as we see in the next section. The notation allowing for a single technical basis is summarized in Table \ref{table:ThetaNotation}, where we must now give our single technical basis the generic label $Z$ just to populate the superscripts. 

%----------------------------------------------------------------

\subsection{Surplus With a Single Technical Basis}
\label{sec:Precursors}

The main result of this section is the following.

\begin{proposition} \label{prop:Precurose} Let ${\cal B}^Z = (\delta_t,\mu_t)$ be a generic technical basis, and for brevity also omit the $Z$ postscript from $V_t$, $R(t)$ etc.. Then for $t \in (0,n)$:

\begin{eqnarray}
\mbox{\rm (a) } \, d{\Theta}^{Z,\omega}(t) & = & \delta_t \, {\Theta}^{Z,\omega}(t) \, dt - R(t) \, dM(t) \label{eq:PrecursorA} \\
\mbox{\rm (b) } \, d \tilde{\Theta}^{Z,\omega}(t) & = & - v(t) \, R(t) \, dM(t) \label{eq:PrecursorB} \\
\mbox{\rm (c) } d{\Theta}^{Z,Z}(t) & = & \delta_t \, {\Theta}^{Z,Z}(t) \, dt \label{eq:PrecursorC} \\
\mbox{\rm (d) } d \tilde{\Theta}^{Z,Z}(t) & = & 0. \label{eq:PrecursorD}
\end{eqnarray}

\end{proposition}

\noindent {\em Proof}: See Proposition \ref{prop:SmallProp}. \hfill{$\Box$}

\medskip

\medskip

\noindent ${\cal B}^Z$ is a special technical basis so $\tilde{\Theta}^{Z,\omega}(0) = \tilde{\Theta}^{Z,Z}(0) = -V_0^Z = 0$, and the terminal net cashflow $\Delta B(n) = - 1_{\{ N(n)=0 \}} \, ( \bar{S} - V_n^Z ) = 0$ so we have the following.

\begin{corollary} \label{corr:IntegrateSurp1} Either by forming complete differentials in (a) and (c) of Proposition \ref{prop:Precurose}, as in \cite[Section 5A]{norberg1999}, or by integrating (b) and (d) directly, we have for $0 \le t \le n$:

\begin{eqnarray}
\mbox{\rm (a) } \tilde{\Theta}^{Z,\omega}(t) & = & - \int_0^t v(r) \, R^Z(r) \, dM^Z(r) \label{eq:PrecursorCorrA} \\
\mbox{\rm (b) } \tilde{\Theta}^{Z,Z}(t) & = & 0. \label{eq:PrecursorCorrB} 
\end{eqnarray}

\end{corollary}

\hfill{$\Box$}

%-------------------------------------------------------------

\subsection{Comments}
\label{sec:CommentProofPrecurose}

\begin{bajlist}

\item {Part (b) of Proposition \ref{prop:Precurose} implies {Hattendorff's Theorem} \citep{buehlmann1976, ramlau-hansen1988b, norberg1992} namely that the discounted insurance surpluses (or losses) over non-overlapping periods are uncorrelated.}

\item Part (b) of Corollary \ref{corr:IntegrateSurp1} expresses the classical equality of prospective and retrospective policy values, which is a statement about ${\cal F}_0$-measureable quantities, see Section \ref{sec:ModSurp}.

\item No boundary values were involved until Corollary \ref{corr:IntegrateSurp1}.  

\end{bajlist}

%---------------------------------------------------------------
%---------------------------------------------------------------
%-------------------------------------------------------------
%-------------------------------------------------------------
%-------------------------------------------------------------
%----------------------------------------------------------------

\begin{table}
\begin{center}
\caption{\label{table:ThetaNotation2TB} `$\Theta$' notation for stochastic and modeled surplus equations at time $t$, with first-order technical basis ${\cal B}^L$ and second-order (experience) basis ${\cal B}^M$. Value of contractual cashflows is  $X(t)$.}
\small
\begin{tabular}{llll}
& & & \\
Name & Notation & Definition & Meas. \\[1.0ex]
% Realized   & $\Theta^{\omega,\omega}(t)$ & $X(t)$ & ${\cal F}_n$  \\[0.5ex]
Stochastic & $\Theta^{L,\omega}(t)$ & $\Mean_L[ \, X(t) \mid {\cal F}_t \, ]$ & ${\cal F}_t$ \\[0.5ex]
Stochastic discounted & $\tilde{\Theta}^{L,\omega}(t)$ & $v^M(t) \, \Mean_L[ \, X(t) \mid {\cal F}_t \, ]$ & ${\cal F}_t$ \\[0.5ex]
Modeled & ${\Theta}^{L,M}(t)$ & $\Mean_M[ \, \Mean_L[ \, X(t) \mid {\cal F}_t \, ] \mid {\cal F}_0 \, ]$ & ${\cal F}_0$ \\[0.5ex]
Modeled discounted & $\tilde{\Theta}^{L,M}(t)$ & $\Mean_M[ \, v^M(t) \, \Mean_L[ \, X(t) \mid {\cal F}_t \, ] \mid {\cal F}_0 \, ]$ & ${\cal F}_0$.
\end{tabular}
\end{center}
\end{table}

%-------------------------------------------------------------

\section{Two Technical Bases: Scandinavian-style Regulation}
\label{sec:StochAndmodeled}

\subsection{Two Technical Bases: First-order and Second-order}
\label{sec:ScandiTechBases}

Consider the two technical bases used under Scandinavian-style regulation.

\begin{bajlist}

\item The {\em first-order technical basis} denoted by ${\cal B}^L$ is a special valuation basis that is intentionally `safe-side' in both interest and mortality. It defines the contractual rate of premium $P_t$ by the equivalence principle, see Section \ref{sec:ThieleSolutions}, so by definition $V_0^L=0$. 

\item The {\em second-order technical basis} is the experience  technical basis ${\cal B}^M$ (Section \ref{sec:TechBasisBM}). Since past cashflows are taken (simplistically) to be accounting facts, this corresponds to the usual practice now of valuing assets at their market value.

\end{bajlist}

%----------------------------------------------------------------

\subsection{Surplus With Two Technical Bases}
\label{sec:SplitNonContract}

The `$\Theta$' notation with two technical bases is summarized in Table \ref{table:ThetaNotation2TB}. Note that $\tilde{\Theta}^{L,\omega}(t)$  is the surplus denoted by $\Gamma_t$ in \cite{ramlau-hansen1988a}, and with a change of sign is  $\Gamma_t^L$ defined in \cite{hacariz2024}, and by equation (5.4) in \cite{norberg1999}. As examples we show the undiscounted stochastic surplus:

\begin{eqnarray}
\Theta^{L,\omega}(t) & = & \Mean_L[ \, X(t) \mid {\cal F}_t \, ] \\[0.5ex]
& = & \int_0^t \frac{v^M(r)}{v^M(t)} \, dB(r) - \underbrace{\Mean_L \left[ \int_t^n \frac{v^L(r)}{v^L(t)} \, d(-B(r)) \, \Big| \, {\cal F}_t \right]}_{\mbox{\scriptsize $1_{\{ N(t)=0 \}} \, V^L_t$  (see Proposition \ref{prop:IdentityV})}}.\label{eq:StochBalSht1stOrder} 
\end{eqnarray}

\noindent and the discounted modeled surplus:

\begin{eqnarray}
\tilde{\Theta}^{L,M}(t) & = & \Mean_M[ \, v^M(t) \, \Mean_L[ \, X(t) \mid {\cal F}_t \, ] \mid {\cal F}_0 \, ] \\[0.5ex]
& = & \Mean_M \left[ \int_0^t {v^M(r)} \, dB(r) \, \Big| \, {\cal F}_0 \right] - \Mean_M \big[ v^M(t) \, 1_{\{ N(t)=0 \}} \, {V}_t^L \mid {\cal F}_0 \big]. 
\end{eqnarray}

%----------------------------------------------------------------

\begin{figure}
\begin{center}
\includegraphics[scale=0.87]{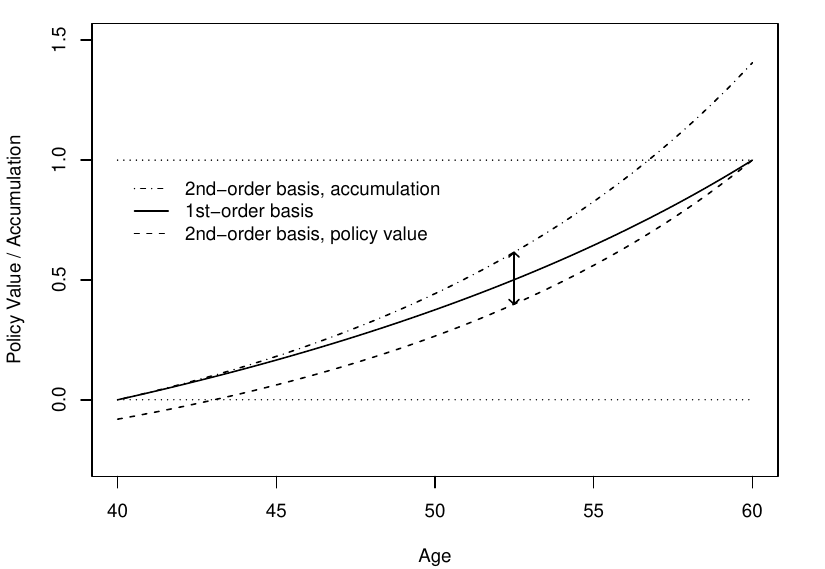}
\caption{\label{fig:PolvalSurplus} Policy values and accumulations on first-order ($\delta^L_t=0.05$, $\mu^L_t$ = GM82 Males (Denmark)) and second-order ($\mu_t^M=0.8 \mu_t^L, \delta_t^M = 1.5 \delta_t^L$) technical bases. Endowment assurance, $S_t = \bar{S} = 1$. Solid line: first-order policy value $V_t^L$. Dashed lines: second-order policy value $V_t^M$ and accumulation $W_t^M$. Vertical line: time-$t$ modeled surplus $\Theta^{M,M}(t)$.}
\end{center}
\end{figure}

%---------------------------------------------------------------

To better understand Scandinavian-style regulation, introduce $V_t^M$, the prospective policy value under ${\cal B}^M$, and surplus $\Theta^{M,\omega}(t)$, \citep{linnemann2002, linnemann2003}:

\begin{eqnarray}
\Theta^{M,\omega}(t) & = & \int_0^t \frac{v^M(r)}{v^M(t)} \, dB(r) - \underbrace{\Mean_M \left[ \int_t^n \frac{v^M(r)}{v^M(t)} \, d(-B)(r) \, \Big| \, {\cal F}_t \right]}_{\mbox{\scriptsize $1_{\{ N(t)=0 \}} \, {V}_t^M$}}. \label{eq:StochBalSht2ndOrder}
\end{eqnarray}

\noindent Then we can interpret the stochastic surplus using the identity:

\begin{equation}
\underbrace{\Theta^{L,\omega}(t)}_{\mbox{\scriptsize Pub. Surp.}} = \underbrace{\Theta^{M,\omega}(t)}_{\mbox{\scriptsize Realistic Surp.}} - \underbrace{1_{\{ N(t)=0 \}} \, ({V}_t^L - {V}_t^M)}_{\mbox{\scriptsize Margin}} \label{BalShtMargin}
\end{equation}

\noindent expressing `published' surplus on the left, as `realistic' surplus less a margin in the liability value. This is illustrated in Figure \ref{fig:PolvalSurplus}, showing policy values and accumulations for a basic endowment contract with unit benefits, on a `safe-side' first-order technical basis ${\cal B}^L$ (solid line) and a `realistic' second-order technical basis ${\cal B}^M$ (dashed lines)\footnote{\cite{linnemann2002, linnemann2003} gave a detailed analysis of surplus based on Thiele's equation (in our terms $\Theta^{L,M}(t)$). All surplus was distributed as terminal bonus, and described as an accumulating bonus account; for example: ``$\ldots$ bonuses are accumulated to the surviving policyholders in the policyholder's account, which is given by Thiele's differential equation $\ldots$'' \citep[p.165]{linnemann2003}. The latter defined $\tilde{B}_r(t)$ to be the {expected accumulation} of past surplus, with $\tilde{B}_r(t)=0$; $\tilde{B}_p(t)$ to be the {EPV} of future surplus, with $\tilde{B}_p(0) = -V_0^M$; and $\tilde{g}(t)$ to be the rate of emerging surplus {($c^{L,M}(t)$ in our terms}, Linnemann's `tilde' does not mean financial discounting). Then (mixing notations slightly) we have the relations:

\begin{eqnarray}
\displaystyle{\frac{d}{dt} \, \tilde{B}_r(t)} & = & (\delta_t^M + \mu^M_t) \, \tilde{B}_r(t) + \tilde{g}(t) \qquad \mbox{\citep[(5.1.3)]{linnemann2003}} \label{eq:Linnemann1} \\
\displaystyle{\frac{d}{dt} \, \tilde{B}_p(t)} & = & (\delta_t^M + \mu^M_t) \, \tilde{B}_p(t) - \tilde{g}(t) \qquad \mbox{\citep[(4.2.5)]{linnemann2003}} \label{eq:Linnemann2}
\end{eqnarray}

\noindent and the whole system is as illustrated in Figure \ref{fig:PolvalSurplus}; initially all future surplus is capitalized in $\tilde{B}_p(0) = -V_0^M$, then as time passes it is transformed into realized surplus {\em via} (\ref{eq:Linnemann1}) and (\ref{eq:Linnemann2}) until finally $\tilde{B}_p(n)=0$, see the comment following Corollary \ref{corr:IntegrateSurp2}.}.

We can now complete the description of Figure \ref{fig:TechBases}. Panel (d) is the same as panel (b), but representing modeled surplus $\Theta^{L,M}(t)$ for $0 \le t < 10$ on two possible technical bases ${\cal B}^M$. This can be regarded as the outcome of scenario testing by an actuary at $t=0$, showing two {possibilities} for ${\cal B}^M$ before $t=10$ and three {choices} of ${\cal B}^L$ after $t=10$.  

%-------------------------------------------------------------

\subsection{Surplus With Two Technical Bases}
\label{sec:SurplusScandi}

Define the {\em systematic component of surplus}, denoted by $c^{L,M}(t)$, {to be the rate}:

\begin{equation}
c^{L,M}(t) = (\delta^M_t - \delta_t^L) \, V_t^L - ( \mu_t^M - \mu_t^L ) \, R^L(t). \label{eq:SystematicComponent}
\end{equation}

\begin{proposition} \label{prop:SmallProp} {For $t \in (0,n)$}:

\begin{eqnarray}
\mbox{\rm (a) } \, d{\Theta}^{L,\omega}(t) & = & \delta_t^M \, {\Theta}^{L,\omega}(t) \, dt + 1_{\{ N(t)=0 \}} \, c^{L,M}(t) \, dt - R^L(t) \, dM^M(t) \label{eq:BigSurpScandiC} \\
\mbox{\rm (b) } \, d \tilde{\Theta}^{L,\omega}(t) & = & 1_{\{ N(t)=0 \}} \, v^M(t) \, c^{L,M}(t) \, dt - v^M(t) \, R^L(t) \, dM^M(t)\label{eq:BigSurpScandiD} \\
\mbox{\rm (c) } d{\Theta}^{L,M}(t) & = & \delta_t^M \, {\Theta}^{L,M}(t) \, dt  + p^M(t) \, c^{L,M}(t) \, dt  \label{eq:BigSurpScandiA} \\
\mbox{\rm (d) } d \tilde{\Theta}^{L,M}(t) & = & \varphi^M(t) \, c^{L,M}(t) \, dt. \label{eq:BigSurpScandiB}
\end{eqnarray}

\end{proposition} 

\noindent {\em Proof}: (a) This proof follows  \cite[Section 5A]{norberg1999}. Differentiate the first term on the right-hand side of equation (\ref{eq:StochBalSht1stOrder}), recalling the definition of $dB(t)$ in equation (\ref{eq:CPRep}) and the identities in equations (\ref{eq:IntegratingFactorZ}), then add/subtract $1_{\{ N(t)=0 \}} \, \delta_t^M \, V^L_t \, dt$ to obtain:

\begin{eqnarray}
&  & \! \! \delta_t^M \, {\Theta}^{L,\omega}(t) \, dt + dB(t) + \delta_t^M \, 1_{\{ N(t)=0 \}} \, V^L_t \, dt \nonumber \\
&  & \quad = \delta_t^M \, {\Theta}^{L,\omega}(t) \, dt + ( 1_{\{ N(t)=0 \}} \, P_t \, dt - S_t \, dN(t)) + 1_{\{ N(t)=0 \}} \, \delta_t^M \, V^L_t \, dt.
\label{eq:Prop1DerivationA}
\end{eqnarray}

\noindent Differentiate the second term on the right-hand side of equation (\ref{eq:StochBalSht1stOrder}):

\begin{eqnarray}
d \big( -1_{\{ N(t)=0 \}} \, V^L_t \big) & = & -1_{\{ N(t)=0 \}} \, dV^L_t - d(1_{\{ N(t)=0 \}}) \, V_t^L \nonumber \\
& = & -1_{\{ N(t)=0 \}} \, \Big( \delta_t^L \, V_t^L + P_t - \mu_t^L \, (S_t - V_t^L) \Big) \, dt + V_t^L \, dN(t). \label{eq:Prop1DerivationB}
\end{eqnarray}

\noindent The last term in equation (\ref{eq:Prop1DerivationA})  and the first term in equation (\ref{eq:Prop1DerivationB}) give:

\begin{equation}
1_{\{ N(t)=0 \}} \, (\delta_t^M - \delta_t^L) \, V_t^L \, dt \label{eq:Prop1DerivationC}
\end{equation}

\noindent while adding/subtracting $1_{\{ N(t)=0 \}} \, \mu_t^M \, (S_t-V_t^L) \, dt$ completes the remaining terms:

\begin{equation}
1_{\{ N(t)=0 \}} \, c^{L,M}(t) \, dt - (S_t - V^L_t) \, dM^M(t) \nonumber
\end{equation}

\noindent in equation (\ref{eq:BigSurpScandiC}).

\smallskip

\noindent (b) Follows immediately from (a) and $d \tilde{\Theta}^{L,\omega}(t) = d ( v^M(t) \, {\Theta}^{L,\omega}(t) )$.

\smallskip

\noindent (c) Take conditional expectations $\Mean_M[ \, \bullet \mid {\cal F}_0 \,]$ of both sides in (a). On the left-hand side, by definition, $\Mean_M[ \, \Theta^{L,\omega}(t) \mid {\cal F}_0 \,] = \Theta^{L,M}(t)$, and on the right-hand side the martingale term vanishes, leaving $1_{\{ N(t)=0 \}}$ as the only stochastic term, with $\Mean_M[ \, 1_{\{ N(t)=0 \}} \mid {\cal F}_0 \, ] = p_t^M$. 

\smallskip 

\noindent (d) Similarly to (c), take conditional expectations of both sides of (b) above. \hfill{$\Box$}

\medskip

{Noting the initial conditions $\tilde{\Theta}^{L,\omega}(0) = \tilde{\Theta}^{L,M}(0) = -V_0^L$, and the terminal net cashflow $\Delta B(n) = - 1_{\{ N(n)=0 \}} \, ( \bar{S} - V_n^L )= 0$ we have the following}.

\begin{corollary} \label{corr:IntegrateSurp2} Either by forming complete differentials in (a) and (c) of Proposition \ref{prop:SmallProp}, or by integrating (b) and (d) directly, we have for $0 \le t \le n$:

\begin{eqnarray}
\mbox{\rm (a) } \tilde{\Theta}^{L,\omega}(t) & = & {-V_0^L} + \int_0^t v^M(r) \, 1_{\{ N(r)=0 \}} \, c^{L,M}(r) \, dr - \int_0^t v^M(r) \, R^L(r) \, dM^M(r) \nonumber \\
&   & \label{eq:SmallPropCorrA} \\
\mbox{\rm (b) } \tilde{\Theta}^{L,M}(t) & = & {-V_0^L} + \int_0^t \varphi^M(r) \, c^{L,M}(r) \, dr. \label{eq:SmallPropCorrB} 
\end{eqnarray}

\hfill{$\Box$} 

\end{corollary}

\noindent {In particular, if the premium rate $P_t$ is fixed as that satisfying the equivalence principle under ${\cal B}^L$, but we calculate policy values under ${\cal B}^M$ in Corollary \ref{corr:IntegrateSurp2}, we have $\tilde{\Theta}^{M,M}(t) = -V_0^M$ (that is, a constant), see equation (\ref{BalShtMargin}) and Figure \ref{fig:PolvalSurplus}.}

%---------------------------------------------------------------

{  % DON'T KNOW WHERE THE CLOSING BRACE IS

\newpage

\subsection{Comments on Proposition \ref{prop:SmallProp}}
\label{sec:SystematicSurplus}

\begin{bajlist}

\item In general, as we see later, we have {cumulative} surplus of the form $C(t) = \int_0^t dC(r)$ and {accumulated} surplus of the form $\Theta(t) = \int_0^t v(r) v^{-1}(t) \, dC(r)$.

\item Clearly Proposition \ref{prop:Precurose} is a special case with $c^{Z,Z}(t)=0$ identically.

\item Reduction or simplification of the stochastic surplus by taking ${\cal F}_0$-conditional expectations  \citep{norberg1999, norberg2001} represents risk management by the law of large numbers, appropriate for an insurer. \cite{moeller2007} and \cite{asmussen2020} propose instead just dropping the martingale terms. This may represent assumption of the individual stochastic risk by another party, such as a reinsurer. 

\end{bajlist}

%---------------------------------------------------------------

\subsection{Origins and Development}
\label{sec:CommentOrigins2TechBasis}

\cite{ramlau-hansen1991} introduced two technical bases in the Markov multiple-state model with constant interest.  Prospective and retrospective policy values satisfying Thiele's equations were assumed under both technical bases. His equations (4.6) and the first part of equation (4.14) are respectively more general versions of Corollary \ref{corr:IntegrateSurp2}(b) and (d) above (allowing for statewise decomposition into components).

%----------------------------------------------------------------

\begin{figure}
\begin{center}
\begin{picture}(150,70)
\put(5,47.5){\dashbox(70,13.5){}}
\put(55,16.0){\dashbox(20,12.5){}}
\put(103,16.0){\dashbox(22,45.0){}}
\put(2.5,14.0){\dashbox(143,49.0){}}
% \put(50,30){\framebox(30,15){}}
% \put(0,15){\framebox(30,15){}}
\put(30,52.5){\vector(1,0){20}}
\put(80,52.5){\vector(1,0){20}}
\put(80,22.5){\vector(1,0){20}}
\put(65,45){\vector(0,-1){15}}
\put(115,45){\vector(0,-1){15}}
\put(15,52.5){\makebox(0,0)[c]{$\Theta^{\omega,\omega}(t)$}}
\put(65,52.5){\makebox(0,0)[c]{$\Theta^{L,\omega}(t)$}}
\put(115,52.5){\makebox(0,0)[c]{$\Theta^{L,M}(t)$}}
\put(65,22.5){\makebox(0,0)[c]{$\tilde{\Theta}^{L,\omega}(t)$}}
\put(115,22.5){\makebox(0,0)[c]{$\tilde{\Theta}^{L,M}(t)$}}
\put(40,56){\makebox(0,0)[c]{\scriptsize $\Mean_L[ \, \bullet \mid {\cal F}_t \, ]$}}
\put(90,56){\makebox(0,0)[c]{\scriptsize $\Mean_M[ \, \bullet \mid {\cal F}_0 \, ]$}}
\put(90,26){\makebox(0,0)[c]{\scriptsize $\Mean_M[ \, \bullet \mid {\cal F}_0 \, ]$}}
\put(59,37.5){\makebox(0,0)[c]{\scriptsize $v^M(t)$}}
\put(109,37.5){\makebox(0,0)[c]{\scriptsize $v^M(t)$}}
\put(40,44){\makebox(0,0)[c]{\scriptsize \cite{norberg1991}}}
\put(41,24.5){\makebox(0,0)[c]{\scriptsize Ramlau-Hansen}}
\put(41,20.5){\makebox(0,0)[c]{\scriptsize (1988a) $\Gamma_t$}}
\put(135,44.5){\makebox(0,0)[c]{\scriptsize Classical}}
\put(135,40.5){\makebox(0,0)[c]{\scriptsize L.I. Maths}}
\put(135,34.5){\makebox(0,0)[c]{\scriptsize Linnemann}}
\put(135,30.5){\makebox(0,0)[c]{\scriptsize (2002, 2003)}}
\put(75,11.5){\makebox(0,0)[c]{\scriptsize \cite{norberg1999, norberg2001, jetses2022}}}
\put(66,5.5){\makebox(0,0)[c]{$\underbrace{\phantom{xxxxxxxxxx}}_{\mbox{\scriptsize ${\cal F}_n$-measureable}} \phantom{xxxxxxxxm} \underbrace{\phantom{xxxxxxxxxx}}_{\mbox{\scriptsize ${\cal F}_t$-measureable}} \phantom{xxxxxxxxxm} \underbrace{\phantom{xxxxxxxxxx}}_{\mbox{\scriptsize ${\cal F}_0$-measureable}}$}}
\put(66,-1.5){\makebox(0,0)[c]{$\underbrace{\phantom{xxxxxxxxxxxxxxxxxxxxxxxxxxxxxxx}}_{\mbox{\scriptsize Stochastic Processes}} \phantom{xxxxxxxxm} \underbrace{\phantom{xxxxxxxxxxx}}_{\mbox{\scriptsize ODEs}}$}}
\end{picture}
\end{center}
\caption{\label{fig:History} Schematic representation {in our notation} of the development of stochastic process life insurance models,  post-\cite{hoem1969, hoem1988}, based on two technical bases and Scandinavian-style regulation. Arrows indicate logical not historical order. `Classical L.I. Maths' refers to deterministic models, models based on the random lifetime $T$ (Section \ref{sec:LifeHistories}), and {ordinary differential equations (ODEs), particularly Thiele's equation}. The `classical' models are shown as derived from more fundamental stochastic process models.}
\end{figure}
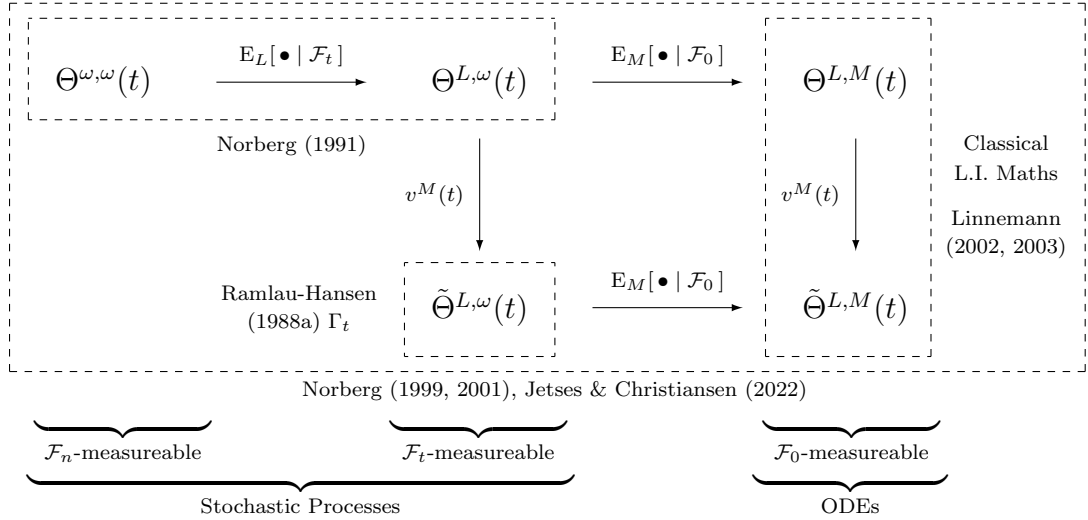

%----------------------------------------------------------------

Proposition \ref{prop:SmallProp} is assembled from materials in  \cite{norberg1996c, norberg1999}. {Part (a) is the same as equation (5.3) in the latter, and part (c) is equivalent to (5.7) (but we do not assume $\tilde{\Theta}^{L,M}(0)=0$). Norberg} formalized parts (c) and (d) of the proposition, essentially by calling upon the law of large numbers. {As noted in Section \ref{sec:SystematicSurplus}(b), approaches to risk management other than the law of large numbers may lead to an alternative simplification of the stochastic surplus.}

{Figure \ref{fig:History} gives a graphical overview of the evolution of stochastic process models of life insurance reserves and surplus, based on Scandinavian-style regulation, after \cite{hoem1969, hoem1988}, and intended to provide context for the $\Theta/\tilde{\Theta}$ notation used here.} { In passing we note the curious fact that in this literature only \cite{width1986} and \cite{linnemann1994, linnemann2002} offer any helpful diagrams other than box-and-arrow depictions of multiple-state models.}

%-------------------------------------------------------------
%-------------------------------------------------------------
%-------------------------------------------------------------

\section{Three Technical Bases: Valuation Premiums and Loadings}
\label{sec:DiffPremValBases}

\subsection{Three Technical Bases}
\label{sec:ThreeBases}

Without going into historical reasons for separating premium and valuation bases (see for example \cite{cox1962, turnbull2017}) we may assume that premium-setting and valuation are seen as having different purposes, especially when the premiums were set a long time ago. Introducing a third technical basis, denoted by ${\cal B}^P$, for the calculation of premiums means that the equation of value satisfied by the contractual premium rate $P_t$ is $\Mean_P[ X(0) \mid {\cal F}_0] = 0$.

Surplus (possibly negative) arises from the difference between ${\cal B}^P$ and ${\cal B}^L$. As we show below, some or all of this surplus may be crystallized at outset into an initial surplus, equal to $-V_0^L$ if there is no countervailing adjustment of assets. More generally, each pair of technical bases (premium, valuation and experience) induces a relationship in terms of surplus, discussed in full in \cite{hacariz2024} from which Figure \ref{fig:RelationsBases2} is taken. 

%------------------------------------------------------------

\begin{figure}
\begin{center}
\begin{picture}(100,65)
\put(50,60.5){\makebox(0,0)[c]{PREMIUM}}
\put(50,55.5){\makebox(0,0)[c]{BASIS}}
\put(15,10.5){\makebox(0,0)[c]{VALUATION}}
\put(15,5.5){\makebox(0,0)[c]{BASIS}}
\put(85,10.5){\makebox(0,0)[c]{EXPERIENCE}}
\put(85,5.5){\makebox(0,0)[c]{BASIS}}
\put(32.5,33){\vector(1,1){15}}
\put(32.5,33){\vector(-1,-1){15}}
\put(67.5,33){\vector(1,-1){15}}
\put(67.5,33){\vector(-1,1){15}}
\put(50,8){\vector(1,0){14}}
\put(50,8){\vector(-1,0){14}}
\put(20.0,35){\makebox(0,0)[c]{{\bf R1}: Surplus}}
\put(20.0,31){\makebox(0,0)[c]{$t = 0$}}
\put(79.5,35){\makebox(0,0)[c]{{\bf R3}: Profit}}
\put(80.5,31){\makebox(0,0)[c]{$t = n$}}
\put(50,4){\makebox(0,0)[c]{{\bf R2}: Surplus Rate}}
\put(50,0){\makebox(0,0)[c]{$0 \le t \le n$}}
\end{picture}
\end{center}
\caption{\label{fig:RelationsBases2} Relationships between the premium basis ${\cal B}^P$, valuation (policy value) basis ${\cal B}^L$ and experience basis ${\cal B}^M$. {\bf R1}: The difference between the premium basis and valuation basis is capitalized as surplus at outset. {\bf R2}: The difference between the valuation basis and experience basis determines the amount and timing of surplus emerging during the policy term. {\bf R3}: The difference between the experience basis and the premium basis determines the final profit. (Source: \cite{hacariz2024})}
\end{figure}
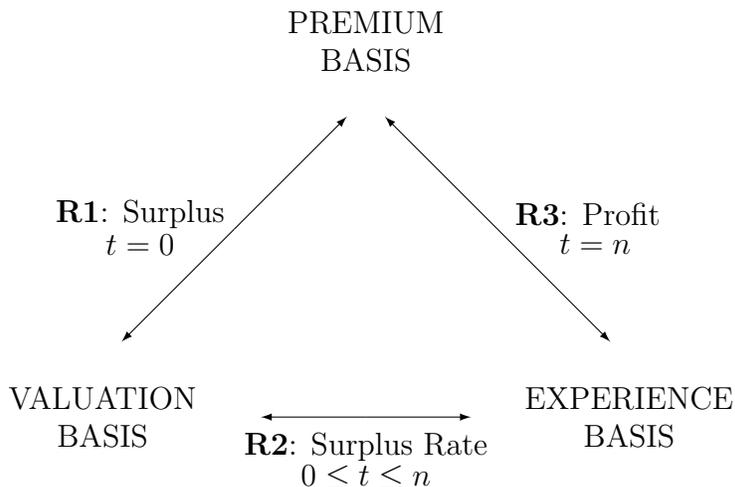

%------------------------------------------------------------

\subsection{The Net (or Pure) Premium Rate}
\label{sec:NetPremium}

Let ${\cal B}^Z$ be a generic technical basis, and let $\pi_t^Z$ be the premium rate satisfying Thiele's equation (\ref{eq:ThieleGeneric}) with boundary conditions $g(0)=0$ and $g(n)=\bar{S}$, under ${\cal B}^Z$. We call $\pi^Z_t$ the  net premium rate  or  pure premium rate for ${\cal B}^Z$ and the contractual benefits. Recall that we assume there to be constraints on the form of the premium rate such that $\pi_t^Z$ is unique, given the contractual benefits, ${\cal B}^Z$ and the boundary condition.

Given separate premium and valuation technical bases ${\cal B}^P$ and ${\cal B}^L$, with pure premium rates $\pi_t^P = P_t$ and $\pi_t^L$ respectively, $(P_t-\pi_t^L)$ is called the {net premium loading} or {pure premium loading}. If $(P_t-\pi_t^L) > 0$ the loading may be regarded as an `extra' rate of premium not required to pay for benefits under ${\cal B}^L$. Therefore, when it is received, it should fall intio surplus, which we call loading surplus. The opposite case $(P_t-\pi_t^L) < 0$, while allowed mathematically, is unsound in practice, because it means that when the contractual premium is received, there must be a charge on surplus in order to make up the required rate of premium $\pi_t^L$. 

%-------------------------------------------------------------

\subsection{Loadings}
\label{sec:InitialSurplusThree}

Since ${\cal B}^P$ and ${\cal B}^L$ are both valuation bases, we denote the corresponding solutions of Thiele's equation by $V_t^P$ and $V_t^L$, satisfying $V_n^P=\bar{S}$ and $V_n^L=\bar{S}$ respectively. By definition, $V_0^P=0$, but in general $V_0^L \not= 0$. In fact we have the following proposition.

\begin{proposition} \label{prop:InitialSurpLoadingsThree}

\begin{equation}
V^L_0 = - \int_0^n {\varphi^L(r)} \, (P_t - \pi_t^L) \, dr.
\end{equation}

\end{proposition}

\noindent {\em Proof}: 

\begin{eqnarray}
V^L_0 & = & \Mean_L \left[ \int_0^n v^L(r) \, \big( d(-B(r)) \, \Big| \, {\cal F}_0 \right] \nonumber \\
& = & \Mean_L \left[ \int_0^n v^L(r) \, \Big( d(-B(r)) + 1_{\{ N(r)=0 \}} \, \big( (P_t-\pi_t^L) - (P_t - \pi_t^L) \big) \, dr \Big) \, \Big| \, {\cal F}_0 \right] \nonumber \\
& = & - \int_0^n \varphi^L(r) \, (P_t - \pi_t^L) \, dr. \label{eq:InitialReserve}
\end{eqnarray}

\hfill{$\Box$}

\noindent It follows that initial surplus is $-V_0^L$, and all the pure premium loadings are capitalized at outset. Inspection of the proof of Proposition \ref{prop:SmallProp} shows that it holds unaltered, since the valuation pure premium rate $\pi_t^L$ is irrelevant.

%-------------------------------------------------------------

\subsection{The Problem of Loadings}
\label{sec:ProblemLoadings}

Separating valuation and premium technical bases creates the pure premium loadings $P_t - \pi_t^L$. These are not  received until the relevant future premiums are paid in future, but they are capitalized and become surplus at outset. This problem preoccupied British actuaries after the Northampton method was abandoned, see \cite{turnbull2017}. In a review of valuation principles, written at about the same time as Thiele formulated his equation, a Scottish actuary said.

\begin{quote} \small ``Alike, in calculating rates of premium and in estimating the value of policies, there are three well-known centres round which all the questions we can consider group themselves --- the table of mortality to be used, the rate of interest to be assumed, and the method of dealing with the `loadings'.'' \citep{mccandlish1876}

\end{quote}

\noindent Note that `loadings', are mentioned, where today we would probably mention `expenses'. Safe-side premiums are larger than they would be based on a `realistic' technical basis. If a valuation then takes a `realistic' view, loadings in future premiums are identified and immediately capitalized as surplus, and if this `surplus' is then disposed of in any form, the very purpose of safe-side premiums would be frustrated. 

The Northampton method did not have this defect, but had defects of its own. Most important  was lack of realism. This became acute in the case of participating business, because premiums were often loaded well beyond mere considerations of risk, with a view to investment. Apart from that, the simple passage of time might render  assumptions made long ago absurd or unsound, especially if they had to be published.

Two basic approaches evolved to `solve' this problem.

\begin{bajlist}

\item For participating business only, to include in the valuation future bonuses sufficient to absorb the future loading surpluses that would otherwise be capitalized. This was called a bonus reserve valuation. It became a valuable tool of internal management, but it was unsuited for use in published valuations because any assumed bonuses could be interpreted as a promise. 

\item To exclude the pure premium  loadings $P_t - \pi_t^L$ from the future premiums valued. That is, in calculating policy values $V_t^L$ by solving Thiele's equation backwards from $V_n^L = \bar{S}$, to substitute the pure premium rate $\pi_t^L$ for the actual premium rate $P_t$. 

\end{bajlist}

Methods along the lines of (a) are called as gross premium valuations, and methods along the lines of (b) are called net premium valuations. The latter became important as the basis of life insurance regulation in many jurisdictions. Their main feature, for our purposes, is that policy values are calculated as if certain non-contractual cashflows were payable, namely the net premium rate $\pi_t^L$ instead of the contractual premium rate $P_t$.

%-------------------------------------------------------------

\subsection{The Valuation Premium Rate}
\label{sec:ValPremRate}

Having raised the possibility of substituting something else for the contractual premium rate $P_t$ in calculating policy values, we realize that the pure premium rate $\pi_t^L$ is not the only possibility. Therefore, let $\tau_t^L$ be {\em any} rate of premium used with valuation technical basis ${\cal B}^L$ and substituted for $P_t$ in calculating $V_t^L$.  We call $\tau_t^L$ a {valuation premium rate}, and we speak of `valuing premium rate $\tau_t^L$'. The two canonical candidates for $\tau_t^L$ we have already discussed: 

\begin{bajlist}

\item if we use $\tau_t^L = P_t$ we have a {gross premium valuation}\footnote{For with-profit business in particular, where premium loadings could include a substantial investment component, this terminology is not primarily related to the inclusion or exclusion of expenses from policy values. By way of contrast see for example \cite[p.175]{hoem1988}: ``At this stage we do not distinguish between net and gross premiums and reserves, i.e. we need not specify whether administrative expenses are taken into account or not.'' This difference in usage is important.};

\item if we use $\tau_t^L = \pi_t^L$ we have a {net premium valuation}; and

\item under Scandinavian-style regulation, the valuation is both net premium and gross premium, since $\tau_t^L = \pi_t^L = P_t$.

\end{bajlist}

\noindent These are not the only possibilities, $\tau_t^L$ may differ from both $P_t$ and $\pi_t^L$. Valuing only some fraction of $P_t$ is one example, the paid-up valuation principle described in Section \ref{sec:ChangingValuationBasis} will provide another, where $\tau_t^L=0$. 

%-------------------------------------------------------------

\subsection{Surplus With Valuation Premium Rate $\tau_t^L$}
\label{sec:InitialSurplus}

We omit the proof of the following proposition, which follows similar lines to the proof of Proposition \ref{prop:InitialSurpLoadingsThree} (see Corollary 5.1 in \cite{hacariz2024}).

\begin{proposition} \label{prop:InitialSurpLoadings} If the valuation premium rate $\tau_t^L$ is substituted for the contractual premium rate $P_t$ in calculating policy values $V_t^L$ under technical basis ${\cal B}^L$, then:

\begin{equation}
V^L_0 = - \int_0^n {\varphi^L(r)} \, (\tau_t^L - \pi_t^L) \, dr.
\end{equation}

\end{proposition}

\hfill{$\Box$}

\noindent That is, $(\tau_t^L - \pi_t^L)$ is a loading which is capitalized at outset. Referring again to Corollary 5.1 in \cite{hacariz2024}, we have the following.

\begin{corollary} \label{corr:Loadings} The pure premium loading $P_t - \pi_t^L$ may be written as:

\begin{equation}
P_t - \pi_t^L = \underbrace{(P_t - \tau_t^L)}_{\mbox{\scriptsize Loading 1}} + \underbrace{(\tau_t^L - \pi_t^L)}_{\mbox{\scriptsize Loading 2}}  \label{eq:PremiumLoadings}
\end{equation}

\noindent in which `Loading 2' is capitalized at once as in Proposition \ref{prop:InitialSurpLoadings} and `Loading 1' falls into surplus when it is received, as we shall see in Proposition \ref{prop:BigProp}. \hfill{$\Box$}

\end{corollary}

Before proceeding to Proposition \ref{prop:BigProp} we need to formalize the use of non-contractual cashflows in calculating policy values and other quantities.

%-------------------------------------------------------------
%-------------------------------------------------------------
%-------------------------------------------------------------

\section{Non-contractual Cashflows and Actuarial Bases}
\label{sec:TechBasisCashflows}

\subsection{Non-contractual Cashflows}
\label{sec:NonConCashflows}

Valuation premium rates $\tau_t^L \not= P_t$ introduce {non-contractual cashflows}. Non-contractual cashflows also arise often other calculations.

\begin{bajlist}

\item {\em Surrender Values}: Premium and valuation bases may commonly omit lapses, equivalent to assuming null lapse rates in a model that otherwise equates the value of benefits on lapsing to the policy value at that time\footnote{Cantelli's Theorem \citep{cantelli1914, hoem1988, christiansen2014}. \cite[Section 8]{hoem1988} noted that second-order calculations might use cashflows (such as surrender values) absent from the corresponding first-order calculations}. However lapseing benefits will certainly appear as an accounting fact in the experience basis and the surplus.

\item {\em Bonus}: {Any distribution of surplus to with-profit policies involves non-contractual cashflows. Bonus is considered in Section \ref{sec:Bonus}.} 

\end{bajlist}

\cite{hacariz2024} defined the `enhanced valuation basis' $\tilde{\cal B}^L$ to be the pair $({\cal B}^L, \tau_t^L)$\footnote{We have changed the notation of \cite{hacariz2024} to agree with ours; the valuation premium rate we call $\tau_t^L$ they called $\pi_t^*$.}. In the following, we make this more general, by parcelling up technical basis and cashflows into a single entity, which we call an {\em actuarial basis}.

%-------------------------------------------------------------

\subsection{Actuarial Bases}
\label{sec:ActBasis}

Suppose we are given a generic technical basis denoted by ${\cal B}^Z$, and, associated with it, some (stochastic) cumulative cashflows denoted by $B^Z(t)$ defining an insurance contract. Define the pair $({\cal B}^Z,B^Z(t))$ to be an {\em actuarial basis}, denoted by ${\cal A}^{Z}$. Actuarial basis ${\cal A}^Z$ parametrizes the Thiele equation:

\begin{equation}
\frac{d}{dt} \, g(t) = \delta_t^Z \, g(t) + \tau_t^Z - \mu_t^Z \, (S_t^Z - g(t)) \label{eq:ThieleActuarialBasis}
\end{equation}

\noindent where $\delta_t^Z$ and $\mu_t^Z$ are from the technical basis ${\cal B}^Z$ and $\tau_t^Z$ and $S_t^Z$ are from the cashflows $B^Z(t)$. If $g(n)=\bar{S}$ the solution is a policy value, ${\cal A}^Z$ is a valuation basis and we may write ${\cal A}^Z = {\cal A}^L$ and $g(t) = V_t^L$, while if $g(0)=0$ the solution is an accumulation and we may write ${\cal A}^Z = {\cal A}^A$ and $g(t) = W_t^A$. 

Table \ref{table:ActuarialBases} shows `standard' definitions. For convenience we denote contractual cashflows equivalently by $B(t) = B^P(t) = B^M(t)$. Note that the valuation actuarial basis ${\cal A}^L$ generalizes and replaces the `enhanced valuation basis' $\tilde{\cal B}^L$ defined in \cite{hacariz2024}.

\begin{table}
\caption{\label{table:ActuarialBases} Characteristics of actuarial bases.}
\small
\begin{center}
\begin{tabular}{lll}
& & \\
Name & Definition & Comment on Cashflow Component \\[0.5ex]
Premium      & ${\cal A}^{P} = ({\cal B}^P , B^P(t))$ & By definition $B^P(t) \equiv B(t)$ \\
Valuation    & ${\cal A}^{L} = ({\cal B}^L , B^L(t))$ & Valuation premium rate $\tau_t^L$ is part of $B^L(t)$ \\
Accumulation & ${\cal A}^{A} = ({\cal B}^A , B^A(t))$ & Default is $B^A(t) = B(t)$, there may be exceptions \\
Experience   & ${\cal A}^{M} = ({\cal B}^M , B^M(t))$ & `True' experience, by definition $B^M(t) \equiv B(t)$. 
\end{tabular}
\end{center}
\end{table}

%-------------------------------------------------------------
%-------------------------------------------------------------
%-------------------------------------------------------------

\section{Surplus with Non-contractual Cashflows}
\label{sec:ModifiedDecomp}

\subsection{Cashflow Values $X^{L,M}(t)$ and Surplus}
\label{sec:SplitNonContractActuarial}

Given the valuation  actuarial basis ${\cal A}^L = ({\cal B}^L,B^L(t))$ and the experience actuarial basis ${\cal A}^M = ({\cal B}^M,B^M(t))$, define the ${\cal F}_n$-measureable $X^{L,M}(t)$ for $0 \le t \le n$ as:

\begin{eqnarray}
X^{L,M}(t) & = & \int_0^t \frac{v^M(r)}{v^M(t)} \, dB^M(r) - \int_t^n \frac{v^L(r)}{v^L(t)} \, d(-B^L)(r). \label{eq:ReproduceSplit3}
\end{eqnarray}

\noindent The valuer looking back {from time $t$ sees {known} cashflows $dB^M(r)$ $(r \le t)$, and looking ahead sees the ${\cal B}^L$-expectation of future cashflows $dB^L(r)$ $(r > t)$}. Then the stochastic surplus at time $t$ is $\Mean_L[ X^{L,M}(t) \mid {\cal F}_t ]$:

\begin{eqnarray}
\Theta^{L,\omega}(t) & = & \Mean_L[ X^{L,M}(t) \mid {\cal F}_t ] \nonumber \\
& = & \int_0^t \frac{v^M(r)}{v^M(t)} \, dB^M(r) - \Mean_L \left[ \int_t^n \frac{v^L(r)}{v^L(t)} \, d(-B)^L(r) \, \Big| \, {\cal F}_t \right]. \label{eq:ReproduceSplit4}
\end{eqnarray}

\noindent Forming the modeled surpluses by taking ${\cal F}_0$-expectations under ${\cal B}^M$, we get (showing only the discounted case as an example):

\begin{eqnarray}
\tilde{\Theta}^{L,M}(t) & = & \Mean_M[ v^M(t) \, \Mean_L[ X^{L,M}(t) \mid {\cal F}_t ] \mid {\cal F}_0 ] \nonumber \\
& = & \Mean_M \left[ v^M(t) \int_0^t \frac{v^M(r)}{v^M(t)} \, dB^M(r) \, \Big| \, {\cal F}_0 \right] \nonumber \\
&   & - \Mean_M \left[ v^M(t) \, \Mean_L \left[ \int_t^n \frac{v^L(r)}{v^L(t)} \, d(-B)^L(r) \, \Big| \, {\cal F}_t \right] \, \Big| \, {\cal F}_0 \right]. \label{eq:ReproduceSplit5}
\end{eqnarray}

\noindent From now on unless stated otherwise, policy values $V_t^L$ and accumulations $W_t^A$, surplus $\Theta^{L,M}(t)$ and $\tilde{\Theta}^{L,M}(t)$, and so on, are based on the referenced {actuarial} bases ${\cal A}^L$ and ${\cal A}^A$ and so on, including their cashflow parts. 

%-------------------------------------------------------------

\subsection{Surplus}
\label{sec:Surplus}

\begin{proposition} \label{prop:BigProp} Let ${\cal A}^M$ and ${\cal A}^L$ be the experience basis and a valuation basis, respectively. Let $R^L(t) = S_t^L - V^L_t$ be the sum at risk at time $t$ under ${\cal A}^L$. Then {for $t \in (0,n)$}:

\begin{eqnarray}
\mbox{\rm (a) } \quad \, {d} {\Theta}^{L,\omega}(t) & = & \delta_t^M \, {\Theta}^{L,\omega}(t) \, dt + 1_{\{ N(t)=0 \}} {c}^{L,M}(t) \, dt \nonumber \\
&   & \quad + d \big( B^M(t) - B^L(t) \big) - R^L(t) \, dM^M(t) \label{eq:BigSurp1c} \\
\mbox{\rm (b) } \quad \, {d} \tilde{\Theta}^{L,\omega}(t) & = & 1_{\{ N(t)=0 \}} \, v^M(t) \, {c}^{L,M}(t) \, dt \nonumber \\
&   & \quad + v^M(t) \, d \big( B^M(t) - B^L(t) \big) - v^M(t) \, R^L(t) \, dM^M(t) \label{eq:BigSurp1d} \\
\mbox{\rm (c) } \quad {d} {\Theta}^{L,M}(t) & = & \delta_t^M \, {\Theta}^{L,M}(t) \, dt + p^M(t) \,  \hat{c}^{L,M}(t) \, dt \label{eq:BigSurp1a} \\
\mbox{\rm (d) } \quad {d} \tilde{\Theta}^{L,M}(t) & = & \varphi^M(t) \, \hat{c}^{L,M}(t) \, dt. \label{eq:BigSurp1b}
\end{eqnarray}

\end{proposition}

\medskip

\noindent {\em Proof}: Follow the same steps as in the proof of Proposition \ref{prop:SmallProp}. At equation (\ref{eq:Prop1DerivationA}), $S_t \, dN(t)$ becomes $S_t^M \, dN(t) = S_t^L \, dN(t) + (S_t^M - S_t^L) \, dN(t)$; and at equation (\ref{eq:Prop1DerivationB}), $P_t$ becomes $\tau_t^L$;  from which after some rearrangement the additional term in $d \big( ( B^M(t) - B^L(t) \big)$ in (a) and (b) arises. Then (c) and (d) follow directly by taking ${\cal F}_0$-conditional expectations of $d \big( ( B^M(t) - B^L(t) \big)$. \hfill{$\Box$}

\bigskip

\noindent Define an augmented systematic part of the surplus\footnote{The {contribution formula} from \cite[p.141]{berger1939} referred to in \cite{ramlau-hansen1988a}  includes the `technical basis component' from equation (\ref{eq:SystematicComponent2}) and a `cashflow component' derived from premium-related and per-policy expenses. The latter component is distinct from the cashflow component in equation (\ref{eq:SystematicComponent2}), which excludes expenses, and in particular {it is not the same as} the valuation premium term $(\tau_t^M - \tau_t^L)$. Likewise, see the contribution formula including expenses in \cite[equation (7.5)]{norberg2001}.} as {the rate}: 

\begin{equation}
\hat{c}^{L,M}(t) = \underbrace{(\delta^M_t - \delta_t^L) \, V_t^L - ( \mu_t^M - \mu_t^L ) \, R^L(t)}_{\mbox{\scriptsize Technical Basis Component}} + \underbrace{(\tau_t^M - \tau_t^L) - \mu_t^M \, (S_t^M - S_t^L)}_{\mbox{\scriptsize Cashflow Component}},  \label{eq:SystematicComponent2}
\end{equation}

\noindent noting that $P_t = \tau_t^M$. Then a simple rearrangement yields the following.

\begin{corollary} \label{corr:BigProp} Under the same assumptions, (a) and (b) of Proposition \ref{prop:BigProp} can be written in terms of systematic parts and martingale residuals as:

\begin{eqnarray}
\mbox{\rm (a) } \quad \, {d} {\Theta}^{L,\omega}(t) & = & 1_{\{ N(t)=0 \}} \hat{c}^{L,M}(t) \, dt + \delta_t^M \, {\Theta}^{L,\omega}(t) \, dt \nonumber \\
&   & \quad - (S_t^M - S_t^L) \, dM^M(t) - R^L(t) \, dM^M(t) \label{eq:CorrBigSurp1c} \\
\mbox{\rm (b) } \quad \, {d} \tilde{\Theta}^{L,\omega}(t) & = & 1_{\{ N(t)=0 \}} \, v^M(t) \, \hat{c}^{L,M}(t) \, dt \nonumber \\
&   & \quad - v^M(t) \, (S_t^M - S_t^L) \, dM^M(t)- v^M(t) \, R^L(t) \, dM^M(t). \label{eq:CorrBigSurp1d}
\end{eqnarray}

\hfill{$\Box$}

\end{corollary}

\medskip

Note the following.

\begin{bajlist} 

\item The augmented systematic component of surplus now includes the loading $(\tau_t^M - \tau_t^L) = (P_t - \tau_t^L)$, see Corollary \ref{corr:Loadings}. Proposition \ref{prop:InitialSurpLoadings} showed valuation premium loadings $(\tau^L - \pi^L)$ to have been capitalized into surplus at outset.

\item The new term $\mu^M_t \, (S_t^M - S_t^L)$ is often zero, which simplifies the results above.  

\item These new terms appear to fall outside the decomposition of surplus into orthogonal martingale-driven components in \cite{jetses2022}. 

\item {Proposition \ref{prop:BigProp} and Corollary \ref{corr:BigProp} express the analysis of surplus well-known in many valuation regimes that admit premium loadings, including {\em inter alia} the treatment of certain expenses, which we have ignored \citep{berger1939}, US-style contribution bonus \citep{homans1863} and approaches that supplanted the Northampton method in the UK to control the emergence of premium loadings into surplus \citep{mccandlish1876,turnbull2017}. They may also be extended to profit-testing techniques that focus entirely on profit loadings \citep{anderson1959}.}

\end{bajlist}

{Noting the initial condition $\tilde{\Theta}^{L,\omega}(0) = \tilde{\Theta}^{L,L}(0) = -V_0^L = 0$, and the terminal net cashflow $\Delta \big( B^M(n) - B^L(n) \big) = - 1_{\{ N(n)=0 \}} \, ( \bar{S}^M - \bar{S}^L )$ we have the following}.

\begin{corollary} \label{corr:IntegrateSurp3} Either by forming complete differentials in (a) and (c) of Proposition \ref{prop:BigProp}, or by integrating (b) and (d) directly, we have for $0 \le t < n$:

\begin{eqnarray}
\mbox{\rm (a) } \tilde{\Theta}^{L,\omega}(t) & = & {-V_0^L} + \int_0^t v^M(r) \, 1_{\{ N(r)=0 \}} \hat{c}^{L,M}(r) \, dr - \int_0^t v^M(r) \, R^L(r) \, dM^M(r) \nonumber \\
&   & \quad - \int_0^t v^M(r) \, \, (S_r^M - S_r^L) \, dM^M(r)  \nonumber \\
&   & \label{eq:BigPropCorrA} \\
\mbox{\rm (b) } \tilde{\Theta}^{L,M}(t) & = & {-V_0^L} + \int_0^t \varphi^M(r) \, \hat{c}^{L,M}(r) \, dr  \label{eq:BigPropCorrB} 
\end{eqnarray}

\noindent and for $t=n$:

\begin{eqnarray}
\mbox{\rm (a*) } \tilde{\Theta}^{L,\omega}(n) & = & {-V_0^L} + \int_0^n v^M(r) \, 1_{\{ N(r)=0 \}} \hat{c}^{L,M}(r) \, dr - \int_0^n v^M(r) \, R^L(r) \, dM^M(r) \nonumber \\
&   & \quad - \int_0^n v^M(r) \, \, (S_r^M - S_r^L) \, dM^M(r) {- v^M(n) \, 1_{\{ N(n)=0 \}} \, ( \bar{S}^M - \bar{S}^L )} \nonumber \\
&   & \label{eq:BigPropCorrA*} \\
\mbox{\rm (b*) } \tilde{\Theta}^{L,M}(n) & = & {-V_0^L} + \int_0^n \varphi^M(r) \, \hat{c}^{L,M}(r) \, dr - {\varphi^M(n) \, (\bar{S}^M - \bar{S}^L)}. \label{eq:BigPropCorrB*} 
\end{eqnarray}

\hfill{$\Box$} 

\end{corollary}

{Corollary \ref{corr:IntegrateSurp3} shows that choosing the valuation basis amounts to choosing a subdivision of the  surplus into surplus capitalized at outset as $-V_0^L$, and surplus that emerges later. Proposition  \ref{prop:InitialSurpLoadings} (and equation (\ref{eq:PremiumLoadings})) shows how this also subdivides the premium loadings. The next section completes the relationships shown in Figure \ref{fig:RelationsBases2} by showing that the discounted modeled surplus $\tilde{\Theta}^{L,M}(n)$ is independent of the valuation basis.}

%-------------------------------------------------------------

\subsection{Invariance Under Change of Valuation Basis}
\label{sec:Invariance}

\cite[Proposition 5.3 (iii)]{hacariz2024} showed that the EPV of total surplus under the experience basis did not depend on the choice of valuation basis\footnote{This result was well-known in jurisdictions where the valuation basis was not mandatory (for example, see \cite{fisher1965}) but had not, to our knowledge, been shown before in the modern setting.}. 

\begin{proposition} \label{prop:Invariance} Under the assumptions of Proposition \ref{prop:BigProp}, the EPV $\tilde{\Theta}^{L,M}(n)$ of the total surplus under the experience basis ${\cal A}^M$ {does not depend on the valuation basis ${\cal A}^L$}.

\end{proposition}

\noindent {\em Proof}: Consider:

\begin{eqnarray}
d \left\{ 1_{\{ N(t)=0 \}} \, v^M(t) \, V_t^L \right\} & = & 1_{\{ N(t)=0 \}} \, d \big( v^M(t) \, V_t^L \big) + d(1_{\{ N(t)=0)\}}) \, v^M(t) \, V_t^L. \label{eq:Invariance1}
\end{eqnarray}

\noindent The first term on the right-hand side gives:

\begin{equation}
v^M(t) \, 1_{\{ N(t)=0 \}} \, \big( - \delta_t^M \, V_t^L +  \delta_t^L \, V_t^L + \tau^L - \mu_t^L \, (S_t^L - V_t^L) \big) \, dt \label{eq:Invariance2}
\end{equation}

\noindent and the second term is $- v^M(t) \, V_t^L \, dN(t)$, see equation (\ref{eq:CPIdentities}). Add and subtract: 

\begin{eqnarray}
v^M(t) \, \big( d(-B)^M(t) + 1_{\{ N(t)=0 \}} \, \mu_t^M \, ({S_t^M} - V_t^L) \, dt \big) & = & v^M(t) \, \Big\{ ( S_t^M \, dN(t) - 1_{\{ N(t)=0 \}} \, P_t \, dt) \nonumber \\
&   & \quad {+ 1_{\{ N(t)=0 \}} \, \mu_t^M \, (S_t^M - S_t^L) \, dt}\nonumber \\
&   & \quad + 1_{\{ N(t)=0 \}} \, \mu_t^M \, (S_t^L - V_t^L) \, dt \Big\} \label{eq:Invariance4}
\end{eqnarray}

\noindent noting that $P_t = \tau_t^M$ and $S_t^M \, dN(t) = S_t^L \, dN(t) + (S_t^M - S_t^L) \, dN(t)$. Gathering terms, the right-hand side of (\ref{eq:Invariance1}) is in total:

\begin{equation}
- 1_{\{ N(t)=0 \}} \, v^M(t) \, \hat{c}^{L,M}(t) \, dt + v^M(t) dB^M(t) + v^M(t) \, ({S_t^M} - V_t^L) \, dM^M(t). \label{eq:Invariance4a}
\end{equation}

\noindent {Then} since:

\begin{equation}
\int_0^n d \left\{ 1_{\{ N(t)=0 \}} \, v^M(t) \, V_t^L \right\} = 1_{\{ N(n)=0 \}} \, v^M(n) \, V_n^L - V_0^L \label{eq:Invariance5}
\end{equation}

\noindent we have, on integrating expression (\ref{eq:Invariance4a}) and equating it to (\ref{eq:Invariance5}):

\begin{eqnarray}
&   & - V_0^L + \int_0^n 1_{\{ N(t)=0 \}} \, v^M(t) \, \hat{c}^{L,M}(t) \, dt \nonumber \\
&   & \quad \quad \quad = - 1_{\{ N(n)=0 \}} \, v^M(n) \, V_n^L + \int_0^n v^M(t) dB^M(t) + \int_0^n v^M(t) \, ({S_t^M}-V_t^L) \, dM^M(t). \nonumber \\
&   & \label{eq:Invariance6}
\end{eqnarray}

\noindent {Add the discounted terminal net cashflow $- v^M(n) \, 1_{\{ N(t)=0 \}} \, (\bar{S}^M - \bar{S}^L)$ to both sides, noting that $V_n^L = \bar{S}^L$}, then taking conditional expectations $\Mean_M[ \, \bullet \mid {\cal F}_0 \, ]$ we have finally:

\begin{eqnarray}
- V_0^L + \int_0^n \varphi^M(t) \, \hat{c}^{L,M}(t) \, dt {- \varphi^M(n) \, (\bar{S}^M - \bar{S}^L)} & = & \Mean_M \left[ \int_0^n v^M(t) dB^M(t) \, \Big| \, {\cal F}_0 \right] {- \varphi^M(n) \, \bar{S}^M} \nonumber \\
& & \label{eq:Invariance7}
\end{eqnarray}

\noindent in which the left-hand side is the EPV under ${\cal B}^M$ of the total surplus with respect to {${\cal A}^L$ {(equation (\ref{eq:BigPropCorrB*}))} and the right-hand side is {the EPV of the total cashflow} and is independent of ${\cal A}^L$}. \hfill{$\Box$}

%-------------------------------------------------------------
%-------------------------------------------------------------
%-------------------------------------------------------------

\section{{Quasi-contractual Cashflows}}
\label{sec:ChangingValuationBasis}

\subsection{{Quasi-contractual Cashflows}}
\label{sec:QuasiContractualCashflows}

No actuarial basis can include all possible events or cashflows. Events beyond the model can change the cashflows that the actuary has to value; we call cashflows subject to such contingencies {quasi-contractual}. The problem they create, in the context of a continuous-time model, is the following. At time $t$, the actuary faces one set of facts and cashflows. By time $t+dt$, the facts and cashflows may have changed. For example:

\begin{bajlist}

\item lapsing a policy triggers payment of any cash value and cancels future premiums; 

\item making a policy paid-up reduces future benefits and cancels future premiums; or

\item declaring a reversionary bonus (an option of the insurer) increases future benefits and (we suppose) leaves future premiums unchanged.

\end{bajlist}

\noindent Examples (a) and (b) above could be handled by transitions to a new state in a multiple-state model. That would not be so easy in example (c). We conclude that there are occasions when it will not do to use the same actuarial basis at time $t+dt$ as was used at time $t$. {Facts} and not only {assumptions} may have changed. 

%----------------------------------------------------------------
 
\subsection{{Continuous Change of Valuation Actuarial Basis}}
\label{sec:ContinuousChange}

We consider the case, which includes the examples above, where the technical basis is fixed but the cashflow component of the valuation actuarial basis changes continuously. We denote this by indexing ${\cal A}^{L_t}$ by time $t$. Heuristically, this means the following.

\begin{bajlist}

\item At time $t$ policy values on $[t,n]$ are calculated using ${\cal A}^{L_t}$. In particular, the policy value at time $t+dt$ is `projected' to be $V_{t+dt}^{L_t}$.

\item By time $t+dt$ the valuation basis has changed to ${\cal A}^{L_{t+dt}}$, and the policy value at time $t+dt$ is actually $V_{t+dt}^{L_{t+dt}}$.

\item Neglecting terms of second order and above, the total increment $dV_t^{L_t}$ is:

\begin{eqnarray}
dV_t^{L_t} & = & \big( V_{t+dt}^{L_{t+dt}} - V_{t}^{L_t} \big) \label{eq:ChangeVa} \\
& = & \big( V_{t+dt}^{L_t} - V_t^{L_t} \big) + \big( V_{t+dt}^{L_{t+dt}} - V_{t+dt}^{L_t} \big). \label{eq:ChangeV}
\end{eqnarray}

\end{bajlist}

\noindent More formally, where Thiele's equation is invoked replace $dV_t^L$ by the total derivative:

\begin{eqnarray} 
dV_r^{L_t} \big|_{r = t} & = & \lim_{r \to t^+} \left( \frac{\partial}{\partial r} \,  V_r^{L_t} \, dr + \frac{\partial}{\partial t} \, V_r^{L_t} \, dt \right) \label{eq:PartialDiffV0} \\
& = & \left[ \big( \delta_t^{L_t} \, V_t^{L_t} + \tau_t^{L_t} - \mu_t^{L_t} \, \left( S_t^{L_t} - V_t^{L_t}  \right) \big) +  \lim_{r \to t^+} \frac{\partial}{\partial t} \, V_r^{L_t} \right] dt \label{eq:PartialDiffV1} 
\end{eqnarray}

\noindent noting that both $r$ and $t$ are annual time units so $dt/dr = 1$.

In the remainder of this section we give two examples involving quasi-contractual cashflows and dynamic actuarial bases: (a) the {paid-up valuation principle} suggested by \cite{linnemann2002, linnemann2003} (Section \ref{sec:PUPValnPrincipleMain}); and (b) some aspects of bonus on participating contracts (Section \ref{sec:Bonus}). It will be helpful in both to follow \cite{linnemann2002} and define the {\em passivum} $k^Z(t)$ at time $t$ under generic {actuarial} basis ${\cal A}^Z$ as:

\begin{equation}
k^Z(t) = \int_t^n \frac{\varphi_r^Z}{\varphi_t^Z} \, \mu_r^Z \, S_r^Z \, dr + \frac{\varphi_n^Z}{\varphi_t^Z} \, \bar{S}^Z  \label{eq:Activum}
\end{equation}

\noindent to be the EPV of future benefit cashflows, and the {\em activum} $a^Z(t)$ as:

\begin{equation}
a^Z(t) = \int_t^n \frac{\varphi_r^Z}{\varphi_t^Z} \, \tau_r^Z \, dr \label{eq:Passivum}
\end{equation}

\noindent to be the EPV of future premium cashflows. So, for example, the equation of value under ${\cal B}^Z$ may be expressed as $k^Z(0) = a^Z(0)$, with $\tau_t^Z$ the unknown premium.

%-------------------------------------------------------------

\begin{figure}
\begin{center}
\includegraphics[scale=0.87]{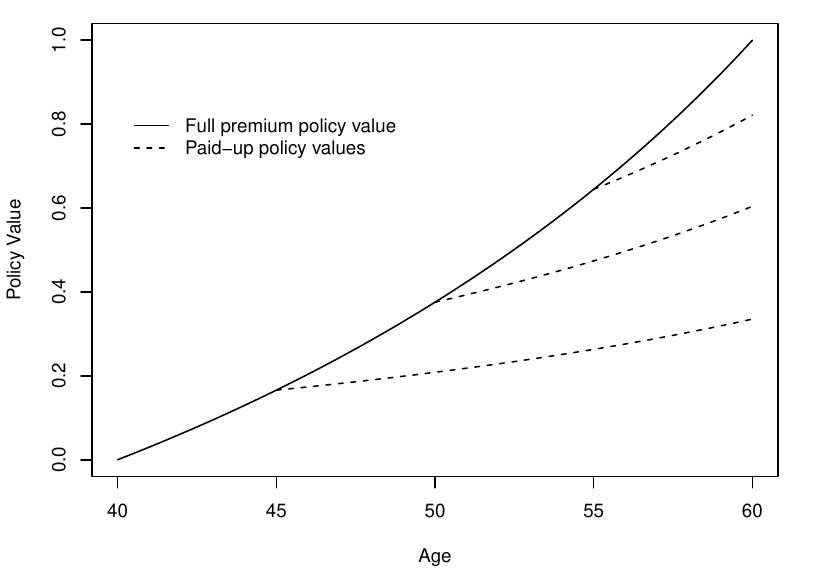}
\caption{\label{fig:PaidUpPolvals} Policy values and paid-up policy values on first-order ($\delta^L_t=0.05$, $\mu^L_t$ = GM82 Males (Denmark)) technical basis. Endowment assurance, $S_t = \bar{S} = 1$. Solid line shows full premium policy values $V_t^L$. Dashed lines show paid-up policy values $V_r^{L_t}$ on $r \in (t,60]$ for policies made paid-up at ages $t = 45, 50$ and 55.}
\end{center}
\end{figure}
 
%----------------------------------------------------------------
%----------------------------------------------------------------
%----------------------------------------------------------------

\section{Example 1: The Paid-up Valuation Principle}
\label{sec:PUPValnPrincipleMain}

\subsection{The Paid-up Valuation Principle}
\label{sec:PUPValnPrinciple}

{Motivated by} Danish regulation, \cite{linnemann2002} introduced a paid-up valuation principle which recognized the policyholder's right at any time to cease paying premiums but leave a reduced sum insured in force. (To be clear, the method values premium-paying policies, but based upon with this right.) The principle assumed that on being made paid-up at time $t$, the prospective policy value $V_t^L$ on the first-order technical basis ${\cal B}^L$ funded reduced benefits for $r \in (t,n]$, which we denote by $S_r^{L_t}$ (death) and $\bar{S}^{L_t}$ (maturity), all {\em pro rata} the original benefits so the `shape' of future benefits was fixed.  

Under this principle, in determining the reserve to be held at time $t$ it is assumed that a premium-paying policyholder makes their policy paid-up at time $t$ and their contractual benefits are reduced by a factor $\kappa^L(t)$ given by:

\begin{equation}
\kappa^L(t) \, k^L(t) = V_t^L. \label{eq:PaidUpCashflows}
\end{equation}

\noindent The cumulative cashflow $B^{L_t}(r)$ under ${\cal A}^{L_t}$ for $t \in (0,n]$ and $r \in (t,n]$ is then:

\begin{equation}
B^{L_t}(r) = \int_0^t \bigg( 1_{\{ N(s)=0 \}} \, \pi^L_s \, ds - S_s^L \, dN(s) \bigg) - \int_t^r \kappa_t^L \, S_s^{L} \, dN(s) - 1_{\{ r=n \}} \kappa_n^L \, \bar{S} \label{eq:PaidUpDefC}
\end{equation}

\noindent and future policy values on $r \in (t,n]$ are given by:

\begin{equation}
V_r^{L_t} = \kappa^L(t) \, k^L(r) = \frac{V_t^L}{k^L(t)} \, k^L(r). \label{eq:PaidUpPolicyValue}
\end{equation}

\noindent Figure \ref{fig:PaidUpPolvals} shows an example of the policy values $V_t^L$ under an endowment assurance (with level unit benefits and level premium rate) compared with the paid-up policy values $V_r^{L_t}$ on $r \in (t,n]$ with the policy being made paid-up at three selected times $t$. 

%-------------------------------------------------------------

\subsection{{Surplus Under the Paid-up Valuation Principle}}
\label{sec:PaidUpSurplus}

Noting that $V_r^{L_t}$ satisfies the Thiele equation:

\begin{equation}
\frac{d}{dr} \, g(r) = \delta^L_r \, g(r) - \mu_r^L \, ( \kappa_t^L \, S_r^L - g(r)) \label{eq:PaidUpThiele}
\end{equation}

\noindent on $r \in (t,n]$ (see Figure \ref{fig:PaidUpPolvals}) and plugging this into equation (\ref{eq:PartialDiffV1}) gives, after some algebra:

\begin{equation}
dV_r^{L_t} \big|_{r = t}  =  \lim_{r \to t^+} \left( \frac{\partial}{\partial r} \,  V_r^{L_t} \, dr + \frac{\partial}{\partial t} \, V_r^{L_t} \, dt \right) = V_t^L.\label{eq:PaidUpPartial_dV}
\end{equation}

\noindent Therefore, valuation on the paid-up principle results in the same policy values and surpluses as before, and all the machinery of Section \ref{sec:Surplus} holds.

Replacing ${\cal A}^M$ by ${\cal A}^L$ in Section \ref{sec:Surplus},  equation (\ref{eq:SystematicComponent2}) gives the systematic surplus:

\begin{eqnarray}
\hat{c}^{L,L_t}(r) \big|_{r = t} & = & d \big( B^L(r) - B^{L_t}(r) \big) \big|_{r = t} \nonumber \\
& = &  \pi_t^L - \mu_t^L \, (S^L_t - S_t^{L_t}). \label{eq:PaidUpSystematic2}
\end{eqnarray}

\noindent Then by setting $r=t$ in equation (\ref{eq:PaidUpPolicyValue}) and differentiating, using (\ref{eq:PaidUpThiele}), we see that:

\begin{equation}
\pi_t^L = \left( \frac{d}{dt} \, \kappa_t^L \right) k^L(t) + \mu_t^L \, (S^L_t - S_t^{L_t}). \label{eq:PaidUpPremiumDecomp}
\end{equation}

\noindent This is the premium decomposition in \cite[Section 3.2]{linnemann2003}:  {if the policyholder chooses to pay the premium $\pi_t^L \, dt$ at time $t$}, it purchases the increment in future benefits and  meets the excess mortality cost (the first and second terms respectively in equation (\ref{eq:PaidUpPremiumDecomp})) leaving net surplus of zero (all under ${\cal A}^L$)

%-------------------------------------------------------------
%-------------------------------------------------------------

\begin{figure}
\begin{center}
\includegraphics[scale=0.87]{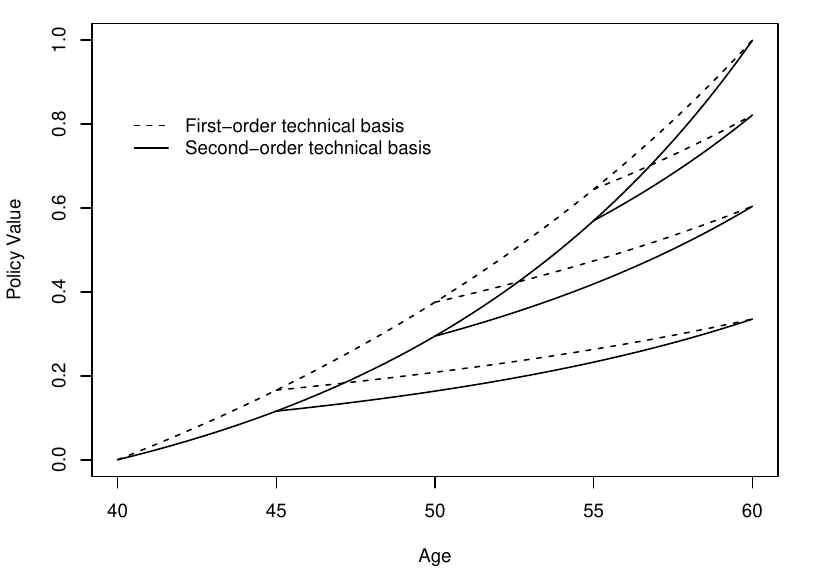}
\caption{\label{fig:PaidUpPolvals2} {Policy values and paid-up policy values on first-order ($\delta^L_t=0.05$, $\mu^L_t$ = GM82 Males (Denmark)) and second-order ($\mu_t^M=0.8 \mu_t^L, \delta_t^M = 1.5 \delta_t^L$) technical bases. Endowment assurance, $S_t = \bar{S} = 1$. Figure shows policy values on second-order basis (solid lines) superimposed on plots from Figure \ref{fig:PaidUpPolvals} (dashed lines).}}
\end{center}
\end{figure}
 
%-------------------------------------------------------------

\subsection{{Paid-up Policy Values Under Alternative Bases}}
\label{sec:PaidUpAltBases}

\cite{linnemann2003} suggested applying the paid-up valuation principle on the second-order technical basis. The actual cashflows are of course unchanged, so the actuarial valuation basis at time $t$, which we may denote by ${\cal A}^{M_t}$, is the pair $( {\cal B}^M , B^{L_t}(r) )$.  Figure \ref{fig:PaidUpPolvals2} shows, as is easily checked, that policy values under ${\cal A}^{M_t}$ are simply those under ${\cal A}^{L_t}$ scaled by a factor $k^M(t)/k^L(t)$. Note that while $V_t^{L_t} = V_t^L$ we have $V_t^{M_t} \not= V_t^M$ (the last quantity as in equation (\ref{eq:StochBalSht2ndOrder})), in fact if ${\cal B}^L$ is a `safe-side' technical basis then:

\begin{equation}
V_t^M \le V_t^{M_t} \le V_t^{L_t} = V_t^L \label{eq:PaidUpInequalities}
\end{equation}

\noindent \cite[equation (6.1.5)]{linnemann2003}, but we do have $V_0^{M_0}=0$. Therefore, this approach allows for some realism of both technical basis and cashflows, reducing capital needs, but without the undesirable feature of capitalizing future surplus at outset.  

Following the same steps as in equation (\ref{eq:PaidUpPremiumDecomp}) we see that at time $t$, a premium of:

\begin{equation}
\pi_t^{M_t} \, dt = \left( \left( \frac{d}{dt} \, \kappa_t^L \right) k^M(t) + \mu_t^M \, (S^L_t - S_t^{L_t}) \right) dt\label{eq:PaidUpPremiumDecompM}
\end{equation}

\noindent will purchase the increment in paid-up benefit and fund the cost of death benefits above $S_t^{L_t}$, so the balance of premium, $\pi_t^L - \pi_t^{M_t}$ emerges as (expected) surplus at time $t$. \cite{linnemann2004} suggested using the approach where: (a) realism is desired in the liability valuation because assets must be taken at market values; and (b) the loadings $\pi_t^L - \pi_t^{M_t}$ in future premiums are bonus loadings which it is desired not to capitalize in advance.

%-------------------------------------------------------------
%-------------------------------------------------------------
%-------------------------------------------------------------

\section{Example 2: Bonus}
\label{sec:Bonus}

\subsection{Surplus, Dividends and Bonus}}
\label{sec:AllocationSurplus}

Under {participating} or {with-profit} policies, surplus is converted into various forms of {bonus} and such an allocation, once {declared}, is irreversible. \cite{ramlau-hansen1991} considered three kinds of bonus, as follows:

\begin{bajlist}

\item {cash} or {contribution bonus} under which surplus is distributed as cash as it is earned;

\item  {reversionary bonus} under which surplus is used to buy additional future benefits; and

\item {terminal bonus} under which surplus is returned to policyholders when they exit.

\end{bajlist}

\noindent We do not have space to consider the merits of bonus systems here, except to note that surplus is an asset, until converted into bonus, at which point it becomes a liability, and the consequent incease in policy values may be called the cost of bonus. The literature is exemplified by \cite[p.184]{asmussen2020} who say: ``The {surplus} $Y$ generates {dividends} that are paid into the {account} $X$ from which the payments, including {bonus} payments, are paid to the policy holders'', a process which is illustrated below.

\begin{equation}
\mbox{Premiums} \; \longrightarrow \; \mbox{Surplus} \; \longrightarrow \; \mbox{Dividends} \; \longrightarrow \; \mbox{Bonus} \; \longrightarrow \; \mbox{Benefits}. \nonumber
\end{equation}

\noindent Mathematical representations must simplify this, see \cite{ramlau-hansen1991, linnemann1994,  norberg1999, norberg2001}. In the example below we omit dividends and model reversionary bonus funded directly out of surplus, which includes methods from UK practice.

%-------------------------------------------------------------

\subsection{An Example: Compound Reversionary Bonus}
\label{sec:BonusValnBases}

{Bonus provides our second example of quasi-contractual cashflows, and} for demonstration we choose reversionary bonus declared continuously, because it leads to transparent procedures. Let ${\cal A}^Z$ be a generic actuarial basis, and associate with it a non-decreasing ${\cal F}_{t^-}$-measureable bonus accumulation function $\tilde{\beta}^Z(t)$ with the following properties.

\begin{bajlist}

\item   The death benefit payable at time $t$ is $S_t^Z \, \tilde{\beta}^Z(t)$, and similarly the maturity value payable at time $n$ is $\bar{S}^Z \, \tilde{\beta}^Z(n)$.

\item There exists an ${\cal F}_{t^-}$-measureable `force of bonus' $\beta^Z_t$ such that $\tilde{\beta}^Z(t) = \exp \big( \int_0^t \beta^Z_r \, dr \big)$ {(therefore $\tilde{\beta}^Z(0)=1$)}. We call $\beta^Z_t$ a {bonus rate} and emphasise that it modifies the cashflows and does not form part of the technical basis ${\cal B}^Z$. 

\end{bajlist}

\noindent {The `null' assumption $\beta^Z_t = 0$ ($t\ge0$) means non-participation, and $\beta^Z_t = b/(1+bt)$ with $b>0$ gives simple reversionary bonus, so the definition is quite general, but clearly it is aligned with assuming $\beta^Z_t$ is constant, which gives level compound reversionary bonus.}

A valuation {actuarial} basis at time $t$ may assume future bonus rates of $\beta^L_r$ $(r \ge t)$, but past bonus declarations at rates $\beta^M_r$ $(r < t)$ are factual. So, we assume that there is a family of valuation actuarial bases $\{ {\cal A}^{L_t} \}_{0 \le t \le n}$, where $B^{L_t}(r)$ is given by:

\begin{equation}
dB^{L_t}(r) = \left\{ \begin{array}{ll} 1_{\{ N(r)=0 \}} \, \tau_r^L \, dt - S_r^M \, \tilde{\beta}^M(r) \, dN(r) & (0 \le r < t) \nonumber \\[1.0ex] \displaystyle{ 1_{\{ N(r)=0 \}} \, \tau_r^L \, dt - \frac{\tilde{\beta}^M(t)}{\tilde{\beta}^{L_t}(t)} \, S_r^L \, \tilde{\beta}^{L_t}(r) \, dN(r)} & (r \ge t) \\[1.5ex] \displaystyle{ -1_{\{ N(n)=0 \}} \, \frac{\tilde{\beta}^M(t)}{\tilde{\beta}^{L_t}(t)} \, \bar{S}^L \, \tilde{\beta}^{L_t}(n)} & (r =n) \label{eq:BonusFactors} \end{array} \right.
\end{equation}

\noindent As a simple example, suppose $\beta^L_t=0.02$ and $\beta^M_t=0.03$, so the valuation actuarial basis anticipates bonus at 2\% {\em per annum} but the experience is that 3\% {\em per annum} is declared. If the basic death benefit is $S$ then policy values calculated at time $t$ anticipate a death benefit of $S \, \exp( 0.03t ) \, \exp( 0.02 ( r - t ))$ at time $r$ ($t \le r \le n$). 

\begin{bajlist}

\item At time $t$, past cashflows $dB^{L_t}(r)$ for $r \in [0,t]$ recognize factual bonus declarations.

\item At time $t$, future cashflows $B^{L_t}(r)$ for $r \in (t,n]$ include assumed bonus declarations after time, part of ${\cal A}^{L_t}$.

\item Technical bases ${\cal B}^{L_t} = {\cal B}^L$ under all ${\cal A}^{L_t}$, hence also expectations $\Mean_{L_t} = \Mean_L$. 

\item The analogue of Scandinavian-style regulation would be assume as the premium basis ${\cal A}^{P}= {\cal A}^{L_0}$, so that $V_0^{L_0}=0$. In the UK this would be called a bonus reserve valuation with premiums calculated on the same basis.

\end{bajlist}

Proceeding to the calculation of surplus, the cashflow value function receives its final embellishment as follows:

\begin{equation}
X^{L_t,M}(t) = \int_{0}^t \frac{v^M(r)}{v^M(t)} \, d{B}^M(r) - \int_{t}^n \frac{v^L(r)}{v^L(t)} \, d(-B^{L_t})(r) \label{eq:CashflowsBonus}
\end{equation}

\noindent and notation for surplus equations allowing for past and future bonus declarations is shown in Table \ref{table:NotationBonus} for convenience. Policy values $V_t^{L_t}$ under ${\cal A}^{L_t}$ are the solution on $(t,n]$ of the Thiele equation:

\begin{equation}
\frac{d}{dr} \, V_r^{L_t} = \delta_r^L \, V_r^{L_t} + \tau_r^{L} - \mu_r^L \, \left( \frac{\tilde{\beta}^M(t)}{\tilde{\beta}^{L_t}(t)} \, S_r^L \, \tilde{\beta}^{L_t}(r) - V_r^{L_t}  \right) \label{eq:ThieleBonus}
\end{equation}

\noindent with boundary condition $V_n^{L_t} = \bar{S}^L \, \tilde{\beta}^{L_t}(n) \, \tilde{\beta}^M(t) \, (\tilde{\beta}^{L_t}(t))^{-1}$. {For brevity we may write $V^L_t$ for $V^{L_0}_t$.} We also need the following definition of the policy value of the benefits alone, where we use the same notation {as the {\em passivum} of equation (\ref{eq:Activum}):

\begin{equation}
k^{L_t,M}(t) = \Mean_L \left[ \frac{\tilde{\beta}^M(t)}{\tilde{\beta}^{L_t}(t)} \, \left( \int_{t}^n \frac{v^L(r)}{v^L(t)} \, S_t^L \, \tilde{\beta}^{L_t}(r) \, dN(r)  + \frac{v^L(n)}{v^L(t)} \, \bar{S}^L \, \tilde{\beta}^{L_t}(n) \, 1_{\{ N(n)=0 \}} \right) \, \Big| \, {\cal F}_t \right]. \label{eq:ThieleBonusSP}
\end{equation}} 

%---------------------------------------------------------------

\begin{table}
\caption{\label{table:NotationBonus} Notation for stochastic and modeled surpluses with cashflows $X^{L_t,M}(t)$ allowing for past and future bonus declarations.}
\small
\begin{center}
\begin{tabular}{lll}
& & \\
Type & Notation & Definition \\[1.0ex]
Stochastic & ${\Theta}^{L_t,\omega}(t)$ & $\Mean_L[ \, X^{L_t,M}(t) \mid {\cal F}_t \, ]$ \nonumber \\[0.5ex] 
Stochastic discounted & $\tilde {\Theta}^{L_t,\omega}(t)$ & $v^M(t) \, \Mean_L[ \, X^{L_t,M}(t) \mid {\cal F}_t \, ]$ \nonumber \\[0.5ex] 
Modeled & ${\Theta}^{L_t,M}(t)$ & $\Mean_M[ \,  \Mean_L[ \, X^{L_t,M}(t) \mid {\cal F}_t \, ] \mid {\cal F}_0 \, ]$ \nonumber \\[0.5ex] 
Modeled discounted & $\tilde{\Theta}^{L_t,M}(t)$ & $\Mean_M[ \, v^M(t) \, \Mean_L[ \, X^{L_t,M}(t) \mid {\cal F}_t \, ] \mid {\cal F}_0 \, ]$.  
\end{tabular}
\end{center}
\end{table}

%-------------------------------------------------------------

The main result of this section is as follows. Define the sums at risk as:

\begin{equation}
R^{L_t}(t) = S_t^L\, \tilde{\beta}^M(t) - V_t^{L_t} \label{eq:SumAtRiskBonus}
\end{equation}

\noindent and the systematic component of surplus as {the rate}:

\begin{eqnarray}
\hat{\hat{c}}^{L_t,M}(t) & = & \underbrace{(\delta^M_t - \delta_t^L) \, V_t^{L_t} - ( \mu_t^M - \mu_t^L ) \, R^{L_t}(t)}_{\mbox{\scriptsize Technical Basis Component}} + \underbrace{(\tau_t^M - \tau_t^L) - \mu_t^M \, (S_t^M - S_t^L) \, \tilde{\beta}^M(t)}_{\mbox{\scriptsize Cashflow Component}} \nonumber \\
&   & \quad - \underbrace{( \beta^M_t - \beta^L_t ) \, k^{L_t,M}(t)}_{\mbox{\scriptsize Cost of Bonus}}. \label{eq:SystematicCompBonus}
\end{eqnarray}

\begin{proposition} \label{prop:BigPropSurplus} Suppose actuarial bases ${\cal A}^M$ (experience) and $\{ {\cal A}^{L_t} \}_{0 \le t \le n}$ (valuation) include in their respective cashflow components reversionary bonus at rates $\beta^M_r$ and $\beta^L_r$ respectively. Then the rate of surplus emerging at time $t$ is:

\begin{eqnarray}
\mbox{\rm (a) } \, \quad d {\Theta}^{L_t,\omega}(t) & = & \delta_t^M \, {\Theta}^{L_t,\omega}(t) \, dt + 1_{\{ N(t)=0 \}} \, \hat{\hat{c}}^{L_t,M}(t) \, dt \nonumber \\
&   & \quad - ( S_t^M - S_t^L) \, \tilde{\beta}^M(t) \, dM^M(t) - R^{L_t}(t) \, dM^M(t) \label{eq:BigSurp1Bonus3} \\
\mbox{\rm (b) } \, \quad d \tilde{\Theta}^{L_t,\omega}(t) & = & 1_{\{ N(t)=0 \}} \, v^M(t) \, \hat{\hat{c}}^{L_t,M}(t) \, dt \nonumber \\
&   & \quad - v^M(t) \, \big( ( S_t^M - S_t^L) \, \tilde{\beta}^M(t) \, dM^M(t) - R^{L_t}(t) \, dM^M(t) \big) \label{eq:BigSurp1Bonus4} \\
\mbox{\rm (c) } \quad d {\Theta}^{L_t,M}(t) & = & \delta_t^M \, {\Theta}^{L_t,M}(t) \, dt + p^M(t) \, \hat{\hat{c}}^{L_t,M}(t) \label{eq:BigSurp1Bonus1} \\
\mbox{\rm (d) } \quad d \tilde{\Theta}^{L_t,M}(t) & = & \varphi^M(t) \, \hat{\hat{c}}^{L_t,M}(t) \, dt. \label{eq:BigSurp1Bonus2} 
\end{eqnarray}

\end{proposition}

\noindent {\em Proof}: (a) Follow the same steps as in Proposition \ref{prop:BigProp} and Corollary \ref{corr:BigProp}. Where $dV_t^L$ appears at equation (\ref{eq:Prop1DerivationB}) (here $dV_t^{L_t}$) differentiate directly {as in equation (\ref{eq:PartialDiffV1})}:

\begin{eqnarray} dV_r^{L_t} & = & \frac{\partial}{\partial r} \,  V_r^{L_t} \, dr + \frac{\partial}{\partial t} \, V_r^{L_t} \, dt \nonumber \\
& = & \left( \delta_r^L \, V_r^{L_t} + \tau_r^{L} - \mu_r^L \, \left( \frac{\tilde{\beta}^M(t)}{\tilde{\beta}^{L_t}(t)} \, S_r^L \, \tilde{\beta}^{L_t}(r) - V_r^{L_t}  \right) \right) \, dr + (\beta_t^M - \beta_t^L) k^{L_t,M}(t) \, dt. \nonumber \\
&   & \label{eq:PartialDiffV} 
\end{eqnarray}

\noindent Both $r$ and $t$ are annual time units so $dt/dr = 1$ and:

\begin{equation}
dV_r^{L_t} \big|_{r = t} = \left( \delta_t^L \, V_t^{L_t} + \tau_t^{L} - \mu_t^L \, R^{L_t}(t) + (\beta_t^M - \beta_t^L) k^{L_t,M}(t) \right) \, dt \label{eq:CompleteDiffV}
\end{equation}

\noindent and the rest of the proof follows the same steps as before.

\medskip

\noindent (b), (c) and (d) follow the same steps as in Proposition \ref{prop:BigProp}, adapted as in (a) above. \hfill{$\Box$}

\medskip

%----------------------------------------------------------------------

\subsection{Comments}
\label{sec:RevBonusComments}

\begin{bajlist}

\item The multiplicative structure of compound bonus additions leads to the appearance of the linear term in $(\beta_t^M - \beta_t^L)$ in the systematic component of the surplus.

\item If $\beta_t^P > 0$, so that future bonus is anticipated in the premium basis, we say the premiums have been explicitly {loaded for bonus}.

\item If $\beta_t^L > 0$, so that future bonus is anticipated in the valuation basis, we have a {bonus reserve} valuation. 

\item {If $\beta_t^M \not= \beta_t^L$ (including a net premium valuation, $\beta_t^L=0$) the systematic component of the surplus includes the additional {cost of bonus} shown in equation (\ref{eq:SystematicCompBonus}).}

\item {Premiums (resp. policy values) may be computed conveniently by calculating the {\em passivum} `as if' for a contract without bonus} at interest rate $\delta_t^P - \beta_t^P$ (resp. $\delta_t^L - \beta_t^L$). This arithmetical sleight of hand has no deeper implications. From there it is a short step to simply using a safe-side technical basis --- Scandinavian-style regulation.

\end{bajlist}

%-------------------------------------------------------------
%-------------------------------------------------------------
%-------------------------------------------------------------
%-------------------------------------------------------------
%-------------------------------------------------------------
%-------------------------------------------------------------
%-------------------------------------------------------------

\newpage

\section{Conclusions}
\label{sec:Concs}

We started by noting the concordance of the Northampton method of policy valuation --- discarded in the UK by about 1850 --- and the Scandinavian-style regulation underlying recent probabilistic models. \cite{norberg2004b} summarized the key features, including: (a) the duality between policy values deterministic as solutions of Thiele's equation; and (b) policy values as conditional expectations in a Markov model. A single first-order technical basis specifying interest and mortality suffices for both. A single technical basis then defines martingale random noise associated with a life insurance contract, see Section \ref{sec:BalanceSheet}  \citep{buehlmann1976, ramlau-hansen1988b}. Two technical bases suffice to describe surplus under Scandinavian-style regulation, see Section \ref{sec:modelingBalanceSheet} \citep{ramlau-hansen1988a, norberg1991}.

Scandinavian-style regulation is characterized by restricting safe-side margins to the elements of the technical basis. Other valuation methodologies admit safe-side loadings also in premiums, and their definition involves the valuation of non-contractual premiums. \cite{hacariz2024} adjusted the valuation technical basis in an {\em ad hoc} manner for this purpose; in Section \ref{sec:TechBasisCashflows} we extended this more generally to an actuarial basis, in which the cashflows as well as the parameters in the associated Thiele equation are specified as part of the basis. This introduces extra elements into the surplus (Section \ref{sec:ModifiedDecomp}). The key outcomes for risk management \citep{hacariz2024} are a split of surplus into loadings capitalized at outset and loadings falling into surplus as premiums are paid (Corollary \ref{corr:Loadings}) and the fact that the conditional expectation $\Mean_M[ \, \bullet \mid {\cal F}_0 \,]$ of surpluses is independent of the valuation actuarial basis (Proposition \ref{prop:Invariance}). 

This analysis can usefully be extended to cases where actual cashflows are dynamic, by introducing a time-indexed family of actuarial bases into the valuation, with a fixed technical basis and variable (quasi-contractual) cashflows in the associated Thiele equations (Section \ref{sec:ChangingValuationBasis}). As examples we applied this construction to surplus arising under the paid-up valuation principle of \cite{linnemann2003} in Section \ref{sec:PUPValnPrincipleMain}, and reversionary bonus in a simple participating setup (bonuses purchased directly out of surplus) in Section \ref{sec:Bonus}.

%-------------------------------------------------------------

\acknowledgements

This study is part of the research programme at the Research Centre for Longevity Risk --- a joint initiative of NN Group and the University of Amsterdam, with additional funding from the Dutch Government's Public Private Partnership programme. We are grateful to Dr Per Linnemann and Prof Dr Marcus Christiansen for comments which helped us to improve the paper.

\competinginterests

None.

%------------------------------------------------------------
%------------------------------------------------------------
%------------------------------------------------------------

\bigskip

\renewcommand{\section}[1]{\noindent}
% \addcontentsline{toc}{section}{References} 
%\bibliography{Bibliography.bib}

%------------------------------------------------------------
%------------------------------------------------------------
%------------------------------------------------------------

\end{document}